\documentclass[aps,twocolumn,pre,superscriptaddress]{revtex4-2}
\usepackage{eurosym}
\usepackage{graphicx}
\usepackage{amsmath}
\usepackage{color}
\usepackage{hyperref}
\hypersetup{
    colorlinks=true,
    linkcolor=blue,
    urlcolor=blue
}

\begin{document}

\title{Solitons in quasiperiodic lattices with fractional diffraction}

\author{Eduard Pavlyshynets}
\affiliation{Department of Physics, \href{https://ror.org/02aaqv166}{Taras Shevchenko National University of Kyiv},
64/13, Volodymyrska Street, Kyiv 01601, Ukraine}

\author{Luca Salasnich}
\affiliation{Dipartimento di Fisica e Astronomia ’Galileo Galilei’ and
Padua QTech Center, \href{https://ror.org/00240q980}{Universit{\`a} di Padova}, via Marzolo 8, 35131 Padova,
Italy}
\affiliation{\href{https://ror.org/00z34yn88}{Istituto Nazionale di Fisica Nucleare (INFN), Sezione di
Padova}, via Marzolo 8, 35131 Padova, Italy}
\affiliation{\href{https://ror.org/02dp3a879}{Istituto Nazionale
di Ottica (INO) del Consiglio Nazionale delle Ricerche (CNR)}, via Nello
Carrara 1, 50019 Sesto Fiorentino, Italy}

\author{Boris A. Malomed}
\affiliation{Department of Physical Electronics, Faculty of Engineering, and Center for Light-Matter Interaction, \href{https://ror.org/04mhzgx49}{Tel
Aviv University}, Tel Aviv 69978, Israel}
\affiliation{Instituto de Alta
Investigaci\'{o}n, Universidad de Tarapac\'{a}, Casilla 7D, Arica, Chile}

\author{Alexander Yakimenko}
\affiliation{Dipartimento di Fisica e Astronomia ’Galileo Galilei’ and
Padua QTech Center, \href{https://ror.org/00240q980}{Universit{\`a} di Padova}, via Marzolo 8, 35131 Padova,
Italy}
\affiliation{Department of Physics, \href{https://ror.org/02aaqv166}{Taras Shevchenko National University of Kyiv},
64/13, Volodymyrska Street, Kyiv 01601, Ukraine}

\begin{abstract}
We study the dynamics of solitons under the action of one-dimensional
quasiperiodic lattice potentials, fractional diffraction, and nonlinearity.
The formation and stability of the solitons are investigated in the
framework of the fractional nonlinear Schr\"{o}dinger equation. By means of
variational and numerical methods, we identify conditions under which stable
solitons emerge, stressing the effect of fractional diffraction on soliton
properties. The reported findings contribute to understanding the soliton
behavior in complex media, with implications for topological photonics and
matter-wave dynamics in lattice potentials.
\end{abstract}

\maketitle

%Fundamental and vortex solitons in %fractional media with\
%quasiperiodic lattice potentials

%Francesco Lorenzi
%Andriy Ronkovych
%Luca Salasnich
%Boris Malomed
%Alexander Yakimenko

\section{Introduction}

%In this study, we explore the intricate %dynamics of solitons in one-dimensional %quasiperiodic lattices under the influence %of fractional diffraction.

%The investigation is driven by the %potential of such phenomena to advance %optical technologies and quantum computing, %where soliton control is crucial.
This work aims to study the interplay between fractional diffraction, which
affects the wave propagation through its nonlocal nature, Anderson
localization (AL), which accounts for the effects of disorder
%and
%quasi-periodicity
on the wave confinement, and local nonlinearity. The
results may help to expand the understanding of the wave dynamics in complex
media, offering insights into new soliton states and stability criteria.

%\emph{What is the physical nature of %fractional diffraction (dispersion)} %\cite{22arxivFrationalDispersionExperiment}.
The fractional diffraction emerges, in the framework of \textit{fractional
quantum mechanic}s \cite{Lask1,Lask2}, as the kinetic-energy operator for
the wave function of particles whose stochastic motion is performed, at the
classical level, by \textit{L\'{e}vy flights} (random leaps). This means
that the average distance $L$ of the randomly
walking classical particle from its initial position grows with time $t$ as
\begin{equation}
L\sim t^{1/\alpha },  \label{eq:DefineLI}
\end{equation}%
where $\alpha $ is the \textit{L\'{e}vy index} (LI) \cite{Mandelbrot}. In
the case of $\alpha =2$, Eq. (\ref{eq:DefineLI}) amounts to the usual random
walk law for a Brownian particle. The L\'{e}vy-flight regime, corresponding
to $\alpha <2$, implies that the corresponding \textit{superdiffusive walk}
is faster than Brownian. The quantization of the L\'{e}vy-flight motion was
performed by means of Feynman's path-integral formulation, in which the
integration is carried out over flight paths characterized by the respective
LI \cite{Lask1,Lask2}. The result is the fractional Schr\"{o}dinger equation (FSE) for wave function $\Psi $ of the L\'{e}vy-flying
particles. In the 1D case, the scaled form of FSE is \cite{Lask1,Lask2,GuoXi}%
\begin{equation}
i\,\frac{\partial \Psi }{\partial t}=\frac{1}{2}\left( -\frac{\partial ^{2}}{%
\partial x^{2}}\right) ^{\alpha /2}\Psi +V(x)\Psi ,  \label{FSE}
\end{equation}%
where $\alpha $ is the same LI as in Eq. (\ref{eq:DefineLI}), and $V(x)$ is
the external potential. The fractional-kinetic-energy
(fractional-diffraction) operator in Eq. (\ref{FSE}) is defined as the
\textit{Riesz derivative}, which is constructed as the juxtaposition of the
direct and inverse Fourier transforms \cite{Riesz},%
\begin{equation}
\left( -\frac{\partial ^{2}}{\partial x^{2}}\right) ^{\alpha /2}\Psi =\frac{1%
}{2\pi }\int\limits_{-\infty }^{+\infty }dp|p|^{\alpha }\int\limits_{-\infty
}^{+\infty }d\xi e^{ip(x-\xi )}\Psi (\xi,t ).  \label{R}
\end{equation}

%%%%%%%%%%%%%%%%%%%%%%%%
%This work discusses 1D solitons. Probably it is better to focus on 1D model.
%%%%%%%%%%%%%%%%%%%%%%%%
%In 2D, FSE takes the form of%
%\begin{equation}
%i\,\frac{\partial \Psi }{\partial %t}=\frac{1}{2}\left( -\frac{\partial ^{2}}{%\partial x^{2}}-\frac{\partial ^{2}}{\partial y^{2}}\right) ^{\alpha %/2}\Psi
%+V(x,y)\Psi ,  \label{FSE2D}
%\end{equation}%
%with the respective operator of the 2D fractional diffraction,%
%\begin{gather}
%\left( -\frac{\partial ^{2}}{\partial x^{2}}-\frac{\partial ^{2}}{\partial
%y^{2}}\right) ^{\alpha /2}\psi =  \notag \\
%\frac{1}{\left( 2\pi \right) ^{2}}\int \int dpdq\left( p^{2}+q^{2}\right)
%^{\alpha /2}\int \int d\xi d\eta e^{i\left[ p(x-\xi )+iq(y-\eta \right]
%}\Psi (\xi ,\eta ).  \label{R2D}
%\end{gather}

{\color{black} This definition of the fractional derivative is a physically relevant one, corresponding to
the intuitively obvious fact that the fractional differentiation of order $\alpha$ amounts to
the multiplication by $|p|^{\alpha}$ in the space of wave number $p$.
Note that this \textit{pseudodifferential} operator is actually nonlocal
for $\alpha\neq 0,2,4, \dots$, as it involves
the wave function values in the whole spatial domain \cite{Jeng_2010_Nonlocality}.
Fractional derivatives naturally extend the concept of AL to regimes
governed by \textcolor{black}{the non-Gaussian statistics}, with L\'{e}vy flights
inducing a heavy-tailed distribution with unique localization
properties}.

The well-known proposal to emulate FSE, which remains far from experimental
realization, by an experimentally accessible equation for paraxial
diffraction of light in an appropriately designed optical cavity \cite{EXP3}%
, and the experimentally realized fractional group-velocity dispersion in a
fiber cavity \cite{Shilong}, suggest a possibility to add the cubic term,
which represents the usual optical nonlinearity, to the respective FSE, thus
arriving at the concept of the fractional nonlinear Schr\"{o}dinger equation
(FNLSE). Models based on diverse varieties of FNLSE are the subject of many
theoretical works \cite{PROP, Chen} aimed to predict fractional
solitons, vortices, domain walls, and other nonlinear modes; see reviews,
Refs. \cite{review,review2}.

%{\color{red}Justify the discussion of %the 2D case (connection with %topological photonics).}

\textcolor{black}{Higher-dimensional solitons with fractional diffraction have potential applications in topological photonics.
In this respect, highly relevant is the two-dimensional (2D) case, as aperiodic lattice potentials offer the basis for creating topological insulators \cite{TI1,TI2}.}
Additional promising possibilities are to
consider the interplay of the fractional diffraction with other forms of
nonlinearity [in particular, nonpolynomial terms \cite{Luca,Delgado}, which
are produced by strong confinement applied to Bose-Einstein condensates (BECs) in the transverse
directions] and, eventually, higher-dimensional settings \cite{book}. In
particular, it may be interesting to construct two- and three-dimensional
localized modes with embedded vorticity.

%{\color{blue} Moreover, aperiodic two-%dimensional lattice potentials,
%cognate to quasiperiodic ones, such as %the Sierpi\'{n}ski gasket, are used as %the
%basis for the creation of topological %photonic setups, such as optical %topological isolators (TIs)
%and higher-order TIs \cite{TI1,TI2}}.

In the framework of the 2D FNLSE,
including self-defocusing nonlinearity {\color{black} (which drives the inner expansion of light beams,
opposite to the shrinkage driven by the self-focusing)} and a spatially periodic potential
(optical lattice, OL, in terms of BECs),
fundamental (zero-vorticity) 2D gap solitons and their stability were
investigated in Refs. \cite{RenDeng} and \cite{Jianhua}. In the latter work,
the case of very deep lattices was addressed, and vortex solitons were
constructed too. In the same model, but with the self-focusing nonlinearity
{\color{black} (which drives shrinkage of light beams due to the Kerr effect)},
vortex solitons of the rhombus and square types (alias onsite- and
offsite-centered ones, {\color{black} with the vortex pivot placed, respectively, at an empty site of the
lattice, or between the sites}) and their stability were addressed in Ref. \cite%
{YaoLiu}. In the case of the usual (nonfractional) diffraction (with LI $%
\alpha =2$) and self-focusing sign of the nonlinearity, rhombus- and
square-shaped vortex solitons, stabilized by the OL potential, %(\ref{V})
were first introduced, respectively, in Refs. \cite{BBB} and \cite{Yang}. In
the limit case of the discrete system with fractional diffraction, vortex
solitons were considered in Ref. \cite{Molina}.
%\section{The problems to be considered}

%In fact, even in the 1D setting with %a 1D
%quasiperidoc potential constructing %stable solitons will be a new problem.
%%%%%%%%%%%%%
%\section{{Solitons in 1D %quasiperiodic lattice with fractional %diffraction }}
%%%%%%%%%%%%%
%\color{black}
%\cbox{pink}{
%I think that the works that considered 1D problems should be noted because that's what we are doing in this %work.
%}

Gap solitons are nonlinear self-trapped modes supported by systems featuring
a bandgap in the linear spectrum \cite{gapsol}. The band gap is a frequency
range in which the periodic structure of the medium suppresses the
propagation of linear waves. The interplay of effects of the OL-induced
band gap and nonlinearity allows the formation of localized wave packets
populating the band gap in the system's spectrum \cite{Brazh,Morsch}. This is
what makes gap solitons special: they exist, and may be stable, in frequency
ranges where the propagation of the linear waves is forbidden by the band gap
(actually, gap solitons cannot represent the system's ground state, i.e.,
they may be metastable modes, at best). Thus, in contrast to the AL, which
is a linear effect, the existence of gap solitons requires the presence of
nonlinearity. Remarkably, the band-gap spectrum of a quasiperiodic potential
is fractal \cite{HS}. The stability and existence of gap solitons in
quasiperiodic lattices with adjustable parameters, such as the sublattice
depth and LI (in the case of the fractional diffraction), have been
investigated in Refs. \cite{HS,Huang_Li_Deng_Dong_2019}. Recently,
localization-delocalization transitions in a 1D linear discrete system
combining fractional diffraction and a quasiperiodic potential were
demonstrated in Ref. \cite{Modak}.

Thus, localized states that are maintained by OLs belong to one of the two
distinct types: (i) low-lying modes of the linear system with a random or
quasiperiodic potential, intrinsically related to the AL; (ii) gap solitons
in nonlinear self-defocusing media, which populate the band gap of the linear
spectrum of the corresponding FSE. \textcolor{black}{Gap solitons with a finite spatial 
extension represent excited states of the nonlinear system that are not produced by 
bifurcations from its linear counterpart. In contrast, AL, being a linear phenomenon, 
is typically suppressed by the self-repulsive nonlinearity \cite{AL_destruction1,AL_destruction2}. 
Here, we focus exclusively on states of the former type, (i).}

{\color{black}
In the present work, we extend the variational method to a multipeak trial wave function, 
building it upon the previously considered single-peak ansatz. This approach allows us to characterize 
the existence domain of the localized states affected by nonlinearity and fractional diffraction. 
The combined effects of the fractional diffraction, self-defocusing, and quasiperiodicity on AL are 
systematically investigated. The stability and symmetry of these states are thoroughly analyzed too.}

The paper is organized as follows. In Sec. \ref{sec:model} we introduce
the model based on the FNLSE with a quasiperiodic potential. In Sec. \ref%
{sec:stationary_states} we present analytical and numerical results for the
stationary states (the analytical results are obtained by means of the
variational approximation, VA). In Sec. \ref{sec:dynamics} we investigate
the stability of stationary states under the action of
self-repulsive nonlinearity. The paper is concluded
in Sec. \ref{sec:conclusions}.

% In the present work, we investigate 1D FNLSE with bichromatic quasiperiodic potential used previously in Ref. \cite{Adhikari_Salasnich_2009}.

% \begin{equation}
% i\frac{\partial \Psi }{\partial t}=\frac{1}{2}\left( -\frac{\partial ^{2}}{%
% \partial x^{2}}\right) ^{\alpha /2}\Psi
% +V(x)\Psi +g \left\vert \Psi \right\vert ^{2}\Psi .  \label{eq:FNLSE1D}
% \end{equation}
% where dimensionless strength of nonlinear interactions corresponds to repulsive nonlinearity ($g>0$)

\section{The model}

\label{sec:model} We aim to generalize, for the case of fractional
diffraction, the model introduced in Ref. \cite{Adhikari_Salasnich_2009} for
BEC in a quasiperiodic potential of the following form: %%%
\begin{equation}
V(x)=\sum_{j=1}^{2}A_{j}\sin ^{2}\left( \frac{2\pi }{\lambda _{j}}\,x\right)
.  \label{eq:V_sin}
\end{equation}%
%
%
%
%
%
%
%%%
Here $A_{j}=4\pi ^{2}s_{j}/\lambda _{j}^{2}$, $s_{j}$ are the amplitudes of
the sublattice potentials, and $\lambda _{j}$ are the respective
wavelengths, measured in units of the oscillatory length $a_{\perp }=\sqrt{%
\hbar /m\omega }$ of the transverse confinement imposed by the tight
harmonic oscillator potential with frequency $\omega $. {\color{black}
%\{1.1, 2.7\}
The \emph{quasiperiodicity} of the potential originates from its
nonperiodic nature, despite the periodicity of the two sublattices
used to create the external trapping potential described by Eq.~(\ref%
{eq:V_sin}). This property arises due to the \emph{incommensurability} of
the sublattice wavelengths, with their ratio $\lambda_2/\lambda_1$ being an
irrational number. The resulting \emph{disorder}, which drives the Anderson
localization of linear waves, is \emph{pseudorandom}, signifying that its
behavior can be analytically predicted.}

A natural conjecture is that an ultracold gas of particles moving by L\'{e}vy flights may form BEC with wave function $\Psi \left( x,t\right) $, that,
in the mean-field approximation, obeys a Gross-Pitaevskii equation built as
FSE (\ref{FSE}), to which the usual collision-induced cubic term is added.
However, it is relevant to mention that a consistent microscopic derivation
of such a fractional Gross-Pitaevskii equation has not yet been reported,
and therefore it may be adopted as a phenomenological model \cite{HS2}.
% {\LARGE [I think this caveat is essential, as no one accurately derived the
% fractional GP equation -- a good subject for another work.]}

In the framework of the conjecture mentioned above, in the effective 1D
setting, the evolution of the wave function $\Psi (x,t)$ of BEC of L\'{e}vy-flying particles is governed by the fractional Gross-Pitaevskii equation,
alias FNLSE. In the scaled form, it is written as
\begin{equation}
i\,\frac{\partial \Psi }{\partial t}=\frac{1}{2}\left( -\frac{\partial ^{2}}{%
\partial x^{2}}\right) ^{\alpha /2}\Psi +V(x)\Psi +g|\Psi |^{2}\Psi ,
\label{eq:GPE}
\end{equation}%
cf. Eq. (\ref{FSE}). The wave function is normalized as follows:
\begin{equation}
\int\limits_{-\infty }^{+\infty }|\Psi (x,t)|^{2}dx=1  \label{eq:N}
\end{equation}%
%
%
%
%
%
%
% We consider the LI range $1<\alpha\le 2$, which is most relevant for various physical realizations \cite{review,review2}.
We here fix the sublattice amplitudes as $s_{1}=2,s_{2}=0.4$ and wavelengths
as
\begin{equation}
\lambda _{1}\equiv \frac{2\pi }{\kappa _{1}}=2,\lambda _{2}\equiv \frac{2\pi
}{\kappa _{2}}=\frac{1}{2}(\sqrt{5}+1)  \label{kappa}
\end{equation}
(i.e., $\lambda _{2}$ is the commonly known golden ratio). %
\textcolor{black}{As mentioned above, in our model disorder arises from the incommensurability
of the two sublattice wavelengths, when the ratio
$\lambda_2/\lambda_1$ is an irrational number.}
% In the present work, we mainly address the case of the repulsive nonlinearity, $g\ge 0$.
% {\LARGE [I have added here the definition of }$\kappa ${\LARGE , as it was not defined -- OK?]}

%%%%%%%%%%%%%%%%%%%%

\section{Stationary states}

\label{sec:stationary_states} %%%%%%%%%%%%%%%%%%%%%

Wave functions of the stationary states are looked for in the usual form,
\begin{equation}
\Psi (x,t)=\psi (x)e^{-i\mu t},
\end{equation}%
where $\mu $ is the chemical potential and the spatial wave function $\psi
(x)$ satisfies the equation
\begin{equation}
\mu \psi =\frac{1}{2}\left( -\frac{\partial ^{2}}{\partial x^{2}}\right)
^{\alpha /2}\psi +V(x)\psi +g|\psi |^{2}\psi .  \label{psi}
\end{equation}%
{\color{black} It is relevant to mention that, even in the absence of the external
potential, the fractional derivative destroys the Galilean invariance of Eq. (%
\ref{eq:GPE}). In fact, this equation gives rise to moving solitons in the
free space ($V(x)=0$), but producing such solutions is a nontrivial problem
\cite{review,review2}.}

Below, we investigate stationary states analytically, by means of VA, and
numerically.
%%%%%%%----------------

\subsection{The variational approach (VA)}

%%%%%%%----------------
\begin{figure}[h]
\includegraphics[width=8.6cm]{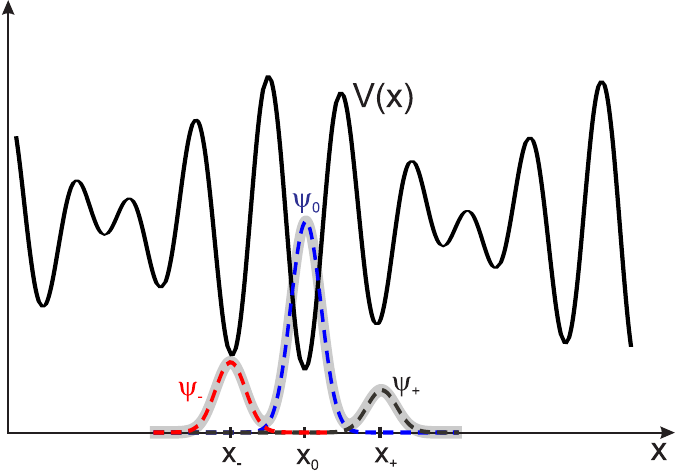}
\caption{Schematics of the quasiperiodic optical lattice potential $V(x)$
(solid black line) and the triple variational ansatz [Eq.~(\protect\ref%
{eq:ansatz})]: the central component $\protect\psi_0(x)$ (dashed blue) and
satellite ones, $\protect\psi_+(x)$ and $\protect\psi_-(x)$ (dashed black
and red lines, respectively). In the general case, satellites exhibit different
peak amplitudes.}
\label{fig:Variat_Potential}
\end{figure}
%{\LARGE [Why the left segment of the %ansatz profile in Fig. 1 is not red %but
%yellowish?]\ }
We have developed VA for the stationary states using a multipeak trial
function (ansatz, see Appendix \ref{sec:general_variational_method}).
{\color{black}The variational solution corresponds to a stationary state
that minimizes the energy functional under the constraint of a fixed norm \eqref{eq:N}.
This method does not necessarily yield the unique ground state but
rather an approximate stationary state that satisfies the energy-minimization principle within the chosen ansatz. It is important to emphasize that a stationary solution, obtained by VA, is not necessarily stable. {\color{black} Stability is determined by the system's response to perturbations rather than the time independence of unperturbed density profile}. In nonlinear wave systems, many stationary solutions, including solitons, can exhibit dynamic instability due to modulational, azimuthal, or collapse-induced effects. Therefore, while the variational method provides an approximate stationary solution, its time evolution is not dictated solely by the method's accuracy but can also reflect the intrinsic instability of the exact stationary solution itself. }

In this section we present VA in detail for a three-peak ansatz of the form
\begin{equation}
\psi (x)=\psi _{0}(x)+\psi _{-}(x)+\psi _{+}(x),  \label{eq:ansatz}
\end{equation}%
where
\begin{equation}
\psi _{j}(x)=h_{j}\exp \left( -\frac{(x-x_{j})^{2}}{2w_{j}^{2}}\right) ,
\label{Gauss}
\end{equation}%
with $h_{j}$ representing amplitudes associated with points $x_{0}$ and $%
x_{\pm }$, that correspond to local minima of the OL (as illustrated in Fig. %
\ref{fig:Variat_Potential}). In ansatz (\ref{eq:ansatz}) the function $\psi
_{0}(x)$ represents the central wave packet, peaking at $x=x_{0}$. We assume
that the effective widths of the satellite components $\psi _{-}(x)$ and $%
\psi _{+}(x)$ are equal to that of the central one, i.e., $%
w_{+}=w_{-}=w_{0}=w$.

% \textcolor{red}{Note that a state localized in a single site is characterized by $h_\pm \ll h_0$, but as the states transition towards a delocalized distribution, satellites emerge in two neighboring sites. Consequently, our generalized variational approach utilizing a multi-peak trial function not only expands the range of applicability for variational results but also provides a quantitative description of the Anderson localization threshold.
% }
% \cbox{pink}{
% We don't see a clear localization threshold from 3-peak variational results (see Fig. \ref{fig:b_pm}). Moreover, I think that any 3-peak distribution can be perceived as localized. I think that delocalization is about spreading of a condensate over many sites, not just 3 or so.
% }

We apply VA only to the overlapping between adjacent peaks, thus focusing on
products of the form $\psi _{0}\psi _{j}$. Terms involving $\psi _{+}\psi
_{-}$ are neglected, as the corresponding overlap integrals are negligibly
small. Therefore, the density distribution and interaction-energy density
are approximated by%
\begin{equation}
|\psi |^{2}\approx \psi _{0}^{2}+\psi _{+}^{2}+\psi _{-}^{2}+2\psi _{0}(\psi
_{+}+\psi _{-}),
\end{equation}
\begin{gather}
|\psi |^{4}\approx \psi _{0}^{4}+\psi _{+}^{4}+\psi _{-}^{4}+6\psi
_{0}^{2}(\psi _{+}^{2}+\psi _{-}^{2}) \\
+4\psi _{0}[\psi _{0}^{2}(\psi _{+}+\psi _{-})+\psi _{+}^{3}+\psi _{-}^{3}].
\end{gather}

In the framework of VA, one obtains the norm (scaled number of particles in
the BEC) as
\begin{equation}
N=\int\limits_{-\infty }^{{+\infty }}|\psi |^{2}dx=\sqrt{\pi }%
wh_{0}^{2}f_{N}(b_{+},b_{-},w),  \label{eq:N_variat}
\end{equation}%
where
\begin{equation}
f_{N}(b_{+},b_{-},w)=1+b_{+}^{2}+b_{-}^{2}+2[\varepsilon
_{+}(w)b_{+}+\varepsilon _{-}(w)b_{-}],
\end{equation}%
\begin{equation}
\varepsilon _{\pm }(w)=\exp \left[ -\frac{(x_{0}-x_{\pm })^{2}}{4w^{2}}%
\right],
\end{equation}%
$b_{\pm }=h_{\pm }/h_{0}$ being ratios of the amplitude of the left and
right satellite peaks to the amplitude of the central one. Utilizing Eq. (%
\ref{eq:N_variat}), the amplitude in the central peak, $h_{0}$, can be
expressed in terms of $N$. Consequently, for a fixed $N$, we have three
variational parameters: the common width of the peaks, $w$, and two relative
amplitudes of the satellites, $b_{\pm }$.

The energy functional for stationary FNLSE (\ref{psi}) is
\begin{gather}
E=\frac{1}{2}\int\limits_{-\infty }^{{+\infty }}dx\,\psi ^{\ast }\left( -%
\frac{\partial ^{2}}{\partial x^{2}}\right) ^{\alpha /2}\psi  \notag \\
+\frac{g}{2}\int\limits_{-\infty }^{{+\infty }}dx\left\vert \psi \right\vert
^{4}+\int\limits_{-\infty }^{{+\infty }}dxV(x)\left\vert \psi \right\vert
^{2}.  \label{eq:Energy}
\end{gather}

Employing ansatz Eq. (\ref{eq:ansatz}), we can analytically calculate all
integrals in Eq. (\ref{eq:Energy}). In particular, the contribution of the
nonlinear self-interaction is linked to the integral
\begin{equation}
\int\limits_{-\infty }^{{+\infty }}|\psi |^{4}dx=\sqrt{\frac{\pi }{2}}%
wh_{0}^{4}f_{g}(b_{+},b_{-},w),
\end{equation}%
where
\begin{gather}
f_{g}(b_{+},b_{-},w)=1+b_{+}^{4}+b_{-}^{4}+6\left( \varepsilon
_{-}^{2}(w)b_{-}^{2}+\varepsilon _{+}^{2}(w)b_{+}^{2}\right)  \notag \\
+4\left( b_{-}(1+b_{-}^{2})\varepsilon
_{-}^{3/2}(w)+b_{+}(1+b_{+}^{2})\varepsilon _{+}^{3/2}(w)\right) .
\end{gather}%
The OL contribution is expressed in terms of the following integral:
\begin{equation}
\int\limits_{-\infty }^{+\infty }\sin ^{2}(\kappa x)|\psi (x)|^{2}dx=\frac{1%
}{2}\sqrt{\pi }wh_{0}^{2}f_{\kappa }(b_{+},b_{-},w),  \label{eq:I_g}
\end{equation}%
where
\begin{gather}
f_{\kappa }(b_{+},b_{-},w)=\sigma _{\kappa }(2\kappa x_{0},w)+\sigma
_{\kappa }(2\kappa x_{+},w)b_{+}^{2}  \notag \\
+\sigma _{\kappa }(2\kappa x_{-},w)b_{-}^{2}+2\varepsilon _{+}(w)\sigma
_{\kappa }(\kappa \lbrack x_{0}+x_{+}],w)b_{+}  \notag \\
+2\varepsilon _{-}(w)\sigma _{\kappa }(\kappa \lbrack x_{0}+x_{-}],w)b_{-},
\end{gather}%
and
\begin{equation}
\sigma _{\kappa }(\xi ,w)=1-e^{-\kappa ^{2}w^{2}}\cos (\xi ).
\end{equation}%
The fractional-diffraction term is given by the integral%
\begin{gather}
\int\limits_{-\infty }^{{+\infty }}dx\,\psi ^{\ast }\left( -\frac{\partial
^{2}}{\partial x^{2}}\right) ^{\alpha /2}\psi =\int\limits_{-\infty }^{{%
+\infty }}dk\,k^{\alpha }\left\vert \mathcal{F}[\psi ]\right\vert ^{2}
\notag \\
=\frac{\sqrt{\pi }h_{0}^{2}}{2w}f_{\alpha }(b_{+},b_{-},w),
\end{gather}%
where $\mathcal{F}$ is the symbol of the Fourier transform, cf. Eq. (\ref{R}%
),
\begin{gather}
f_{\alpha }(b_{+},b_{-},w)=\frac{2}{\sqrt{\pi }}w^{2-\alpha }\Gamma \left(
\frac{1+\alpha }{2}\right) \left\{ 1+b_{+}^{2}+b_{-}^{2}\right.  \notag \\
\left. +2b_{+}\chi _{+}\left( \alpha ,w\right) +2b_{-}\chi _{-}\left( \alpha
,w\right) \right\}  \label{f}
\end{gather}%
and
\begin{equation}
\chi _{\pm }(\alpha ,w)=_{1}F_{1}\left( \frac{1+\alpha }{2},\frac{1}{2},-%
\frac{(x_{0}-x_{\pm })^{2}}{4w^{2}}\right) ,
\end{equation}%
$_{1}F_{1}(a,b,z)$ being the Kummer's confluent hypergeometric function. For
$\alpha =2$, function $f_{\alpha }$ in Eq. (\ref{f}) can be expressed in
terms of elementary functions, using a known property of the hypergeometric
function: $_{1}F_{1}(3/2,1/2,z)=e^{z}(1+2z)$ and $\Gamma (3/2)=\sqrt{\pi }/2$%
.

Finally, we obtain the total VA energy functional:
\begin{gather}
E(b_{+},b_{-},w)=\frac{N}{4w^{2}}I_{\alpha }+\frac{gN^{2}}{2w\sqrt{2\pi }}%
I_{g}  \notag \\
+\frac{N}{2}\left( A_{1}I_{V}^{(1)}+A_{2}I_{V}^{(2)}\right) ,
\label{eq:E_variat}
\end{gather}%
where $I_{\alpha }$, $I_{g}$, and $I_{V}^{(j)}$ are functions of three
variational parameters, $b_{+}$, $b_{-}$, and $w$, defined as follows:
\begin{equation}
I_{\alpha }=f_{\alpha }/f_{N},I_{g}=f_{g}/f_{N}^{2},I_{V}^{(j)}=f_{\kappa
_{j}}/f_{N}.  \label{eq:functionI}
\end{equation}%
In the limit case of the nonfractional diffraction ($\alpha =2$) and for
single-peak ansatze, with $b_{\pm }=0$, the energy functional (\ref{eq:Energy})
carries over into the one obtained by means of VA in Ref. \cite%
{Adhikari_Salasnich_2009}.

Maintaining the normalization condition $N=1$ for fixed LI $\alpha $, and
fixed values of the normalized interaction strength $g$ and OL parameters,
such as the positions of potential-trap minima ($x_{0},x_{\pm }$),
amplitudes ($A_{1},A_{2}$), and $\kappa _{1},\kappa _{2}$ [see Eq. (\ref%
{kappa})], we aim to find values of the variational parameters $%
b_{+},b_{-},w $ that minimize energy $E(b_{+},b_{-},w)$.
{\color{black}To this end, we used the MATLAB standard function "fminsearch" from the Optimization Toolbox, which is based on the Nelder-Mead simplex algorithm \cite{Nelder_Mead_simplex}.}

{\color{black}It is worth noting that although the stationary states are obtained by means of the energy-minimization procedure, it is subject
to restrictions, such as the finite number of degrees of freedom in the trial function (which is equal to the number
of peaks) and specified locations of the peaks. These restrictions put a lower bound on the variational energy, which
remains greater than the true minimum. Thus, the obtained solutions may correspond to metastable states but not to the ground state.}

\begin{figure}[h]
\includegraphics[width=0.48\textwidth]{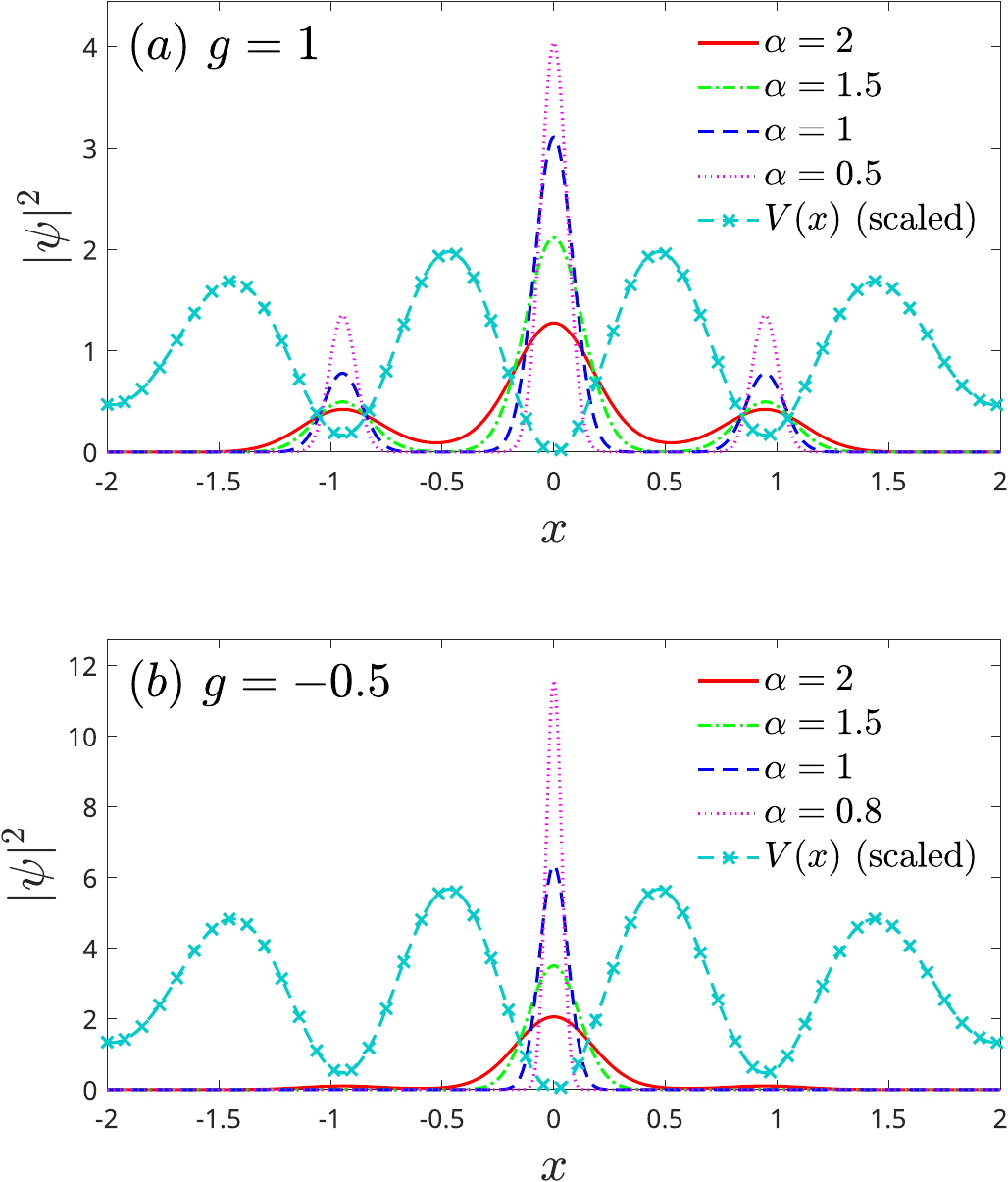}
\caption{The VA solutions for the three-peak ansatz, obtained in the case of
fixed $g=1$ (a) and $g=-0.5$ (b) self-repulsive and self-attractive
nonlinearities, respectively, and different values of LI ($\protect\alpha $).
}
\label{fig:var_alpha}
\end{figure}

Further, the density profiles for fixed values of $g=1$, $g=-0.5$, and
different values of LI $\alpha $ are plotted in Fig. \ref{fig:var_alpha}.
The peaks tend to narrow with the decrease of $\alpha $, which is explained
by the fact that sharper profiles are necessary to balance the nonlinearity
in the case of weaker diffraction. Note that in the cases of $\alpha =1$
and $\alpha <1$ (such as $\alpha =0.8$, shown in Fig. \ref{fig:var_alpha}),
combined with the self-attraction ($g<0$), soliton solutions produced by the
1D FNLSE are unstable, severally, against the action of the critical and
supercritical collapse \cite{review,review2}. {\color{black} The collapse implies the emergence of a
singularity {\color{black}(with the wave function concentrated in an infinitesimally small volume)} 
in the self-focusing medium after a finite propagation distance (the collapse is
critical if the singularity is produced by the input whose power exceeds a finite
critical value; in the case of the supercritical collapse, the critical power vanishes,
i.e., even an arbitrarily weak input gives rise to the collapse)}.
% {\LARGE [I think this caveat,
% concerning the collapse in the case of }$g<0${\LARGE , is essential. Is
% there a chance to add a figure similar to Fig. 3 but with }$g>0${\LARGE ?]}

%To this end, we present the one-%parameter family of VA solutions by
%plotting energy $E$ and $\mu $ %against $g$ in Fig. \ref{fig:mu_E}.

\begin{figure}[h]
\includegraphics[width=0.48\textwidth]
{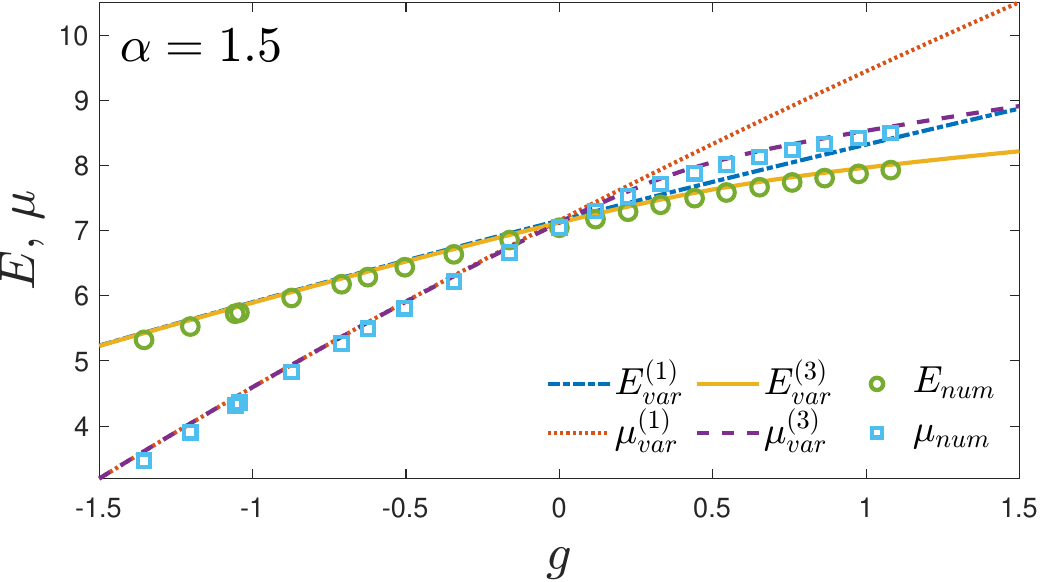}
\caption{Numerical results (circles and squares) and their VA-produced
counterparts, obtained by means of the ansatz admitting the single or three
peaks, for the dependence of energy $E$ and chemical potential $\protect\mu $
on the coupling constant $g$, for LI $\protect\alpha =1.5$. Solid and dashed
lines represent the energy and chemical potential, respectively, of the
three-peak VA solutions, while dash-dotted and dotted lines correspond to
the energy and chemical potential, respectively, of the one-peak VA
solutions.}
\label{fig:mu_E}
\end{figure}

Using the VA solutions, we derive the chemical potential $\mu$ and energy $E$
as functions of the interaction strength $g$ under fixed normalization $N=1$.
Figure \ref{fig:mu_E} illustrates the dependence of the chemical potential
and energy on $g$ for the fixed value of LI, $\alpha=1.5$, comparing results
from VA and numerical methods. The VA based on the three-peak ansatz (\ref%
{eq:ansatz}) yields highly accurate results, while the single-peak ansatz
performs significantly worse.

% {\LARGE %
%[It is stated here that }$\mu (g)$

%{\LARGE \ is displayed in Fig. 5; %however,
%{\Huge no such plots whatsoever} are %present in the figure.]}
% In addition to studying the energy and chemical potential, it is highly informative to visualize the dependence of $b_\pm$ on the interaction strength $g$. This visualization is expected to elucidate the threshold of Anderson localization, particularly as $b_\pm$ begin to increase when self-defocusing nonlinearity $g$ increases, signifying the onset of delocalization.

%{\LARGE [It is insufficient to plot the dependences of %energy and chemical potential on $g$ soley for %$\alpha=1.5$ and 2. Something like $\alpha =1.1$ or 1.2 %would be very relevant to add.]}
\begin{figure}[h]
\includegraphics[width=0.48\textwidth]
{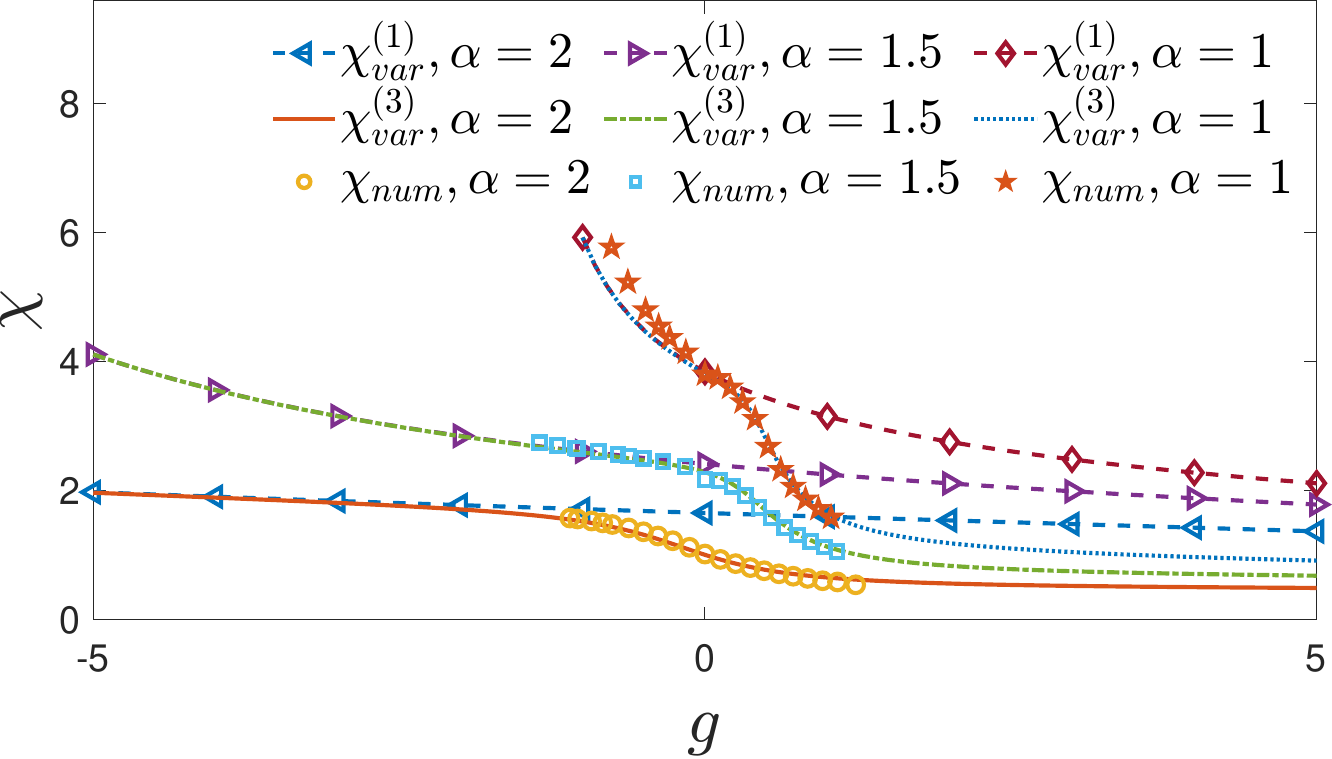}
\caption{Numerical results and their VA-produced counterparts, obtained by
means of the ansatz admitting the single or three peaks, for the dependence
of the form factor [$\protect\chi $, see Eq. \eqref{eq:chi}] on the coupling
constant $g$ for different values of LI and the norm fixed as per Eq. (\ref{eq:N}): $\protect\alpha =2$, $\protect%
\alpha =1.5$, and $\protect\alpha =1$. Note that for $%
\protect\alpha=1$ the curve breaks off in the region of negative $g$,
indicating the onset of the soliton collapse.
{\color{black}Near this region the form factor
diverges, indicating the onset of the singularity, i.e.,
concentration of the wave function in an infinitesimally small volume.}}
\label{fig:chi_plot}
\end{figure}

To characterize the degree of localization of the bound states, we define
the integral form factor $\chi $, fixing the normalization condition %
\eqref{eq:N}:
\begin{equation}
\chi =\int\limits_{-\infty }^{+\infty }|\psi |^{4}dx.  \label{eq:chi}
\end{equation}%
%
%
%
%
% {\LARGE [There is contradiction here with Eq. (\ref{eq:N}) the above statement that the norm is 1.]}
{\color{black} Using the energy functional $E$ from Eq.~(\ref{eq:Energy}), the form factor $\chi$, and the normalization condition \eqref{eq:N}, the chemical potential is given by $\mu = E + \frac{1}{2} {g}\chi$.}

\begin{figure}[h]
\includegraphics[width=0.48\textwidth]
{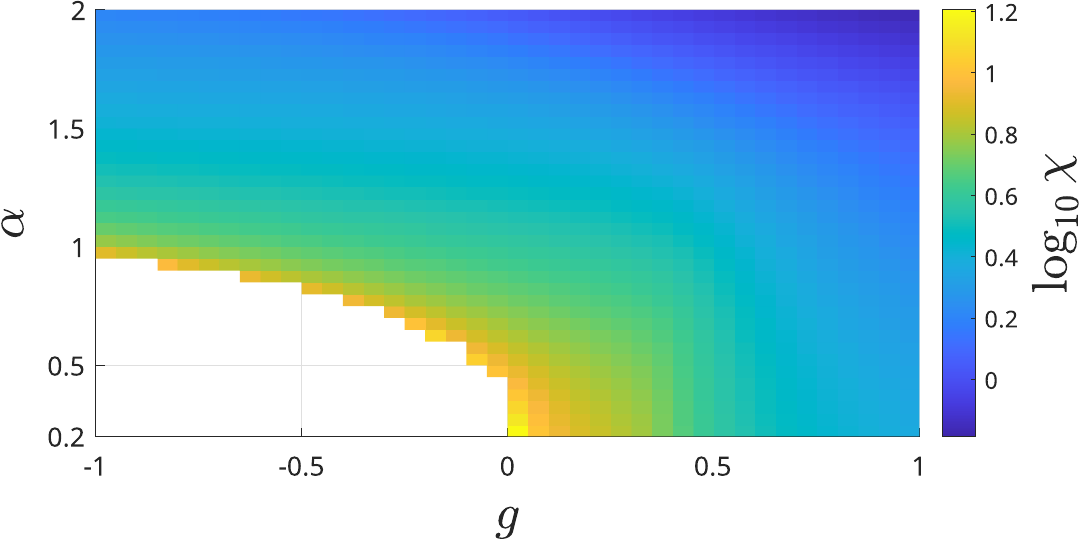}
\caption{VA results for three-peak states, which display the dependence of
the form factor [$\protect\chi $, see Eq. \eqref{eq:chi}] on the coupling
constant $g$ and LI $\protect\alpha $. \textcolor{black}{Note that the VA yields no solutions,
in the white region, for the attractive self-interaction ($g<0$). This happens because the decrease in $\alpha$ gives rise to
the collapse, thereby preventing the existence of solutions in this parameter regime.}}
\label{fig:chi_plot_2D}
\end{figure}

Larger values of the form factor correspond to stronger localization.
% Fig. \ref{fig:chi_plot} shows the comparison of the form-factor dependence on the coupling constant $g$ for different values of the L\'{e}vy index $\alpha$. It can be seen that for every fixed value of the coupling constant $g$ localization is stronger for smaller values of the L\'{e}vy index $\alpha$.
Figure \ref{fig:chi_plot} compares the form factor given by Eq. (\ref{eq:chi}),
obtained using VA and numerical methods, for different values of LI $\alpha$%
. The results produced by three-peak VA for the form factor are summarized
by the heat map in the plane of the nonlinearity strength (coupling constant,
$g$) and LI $\alpha $ in Fig. \ref{fig:chi_plot_2D}.
% {\LARGE \ [I have written in the caption
% to the figure that \textquotedblleft VA produces no solutions in the white
% area." OK?]}

%%%%%%%%
%The main text of the revised manuscript includes now comparison for different values of LI, alpha. We discuss the results for alpha=2 in the Appendix and Supplmental materials.
%%%%%%%
% {\LARGE There are two problems with the presentation of  VA: almost al;l results are plotted solely for the usual diffraction, $\alpha=2$, while it will be more interesting to plot them for $\alpha < 1$, and the ansatz for multi-peak states, with $\geq 5$ peaks, which are presented in the figures, has not been defined. Actually, the single-peak ansatz was not explicitly introduced either. Is it possible to look for dual-peak bound states, with identical or opposite signs of the wave functions at the two peaks?}

\subsection{Two-peak states}

Next, we consider two-peak in-phase and out-of-phase states (with the same
or opposite signs of the peaks of the wave function, respectively). The
locations of the ansatz's peaks are $x_{0}\approx 8$ and $x_{1}\approx 9$,
where the depths of the potential wells take close values.

\begin{figure*}[tbp]
\includegraphics[width=\textwidth]{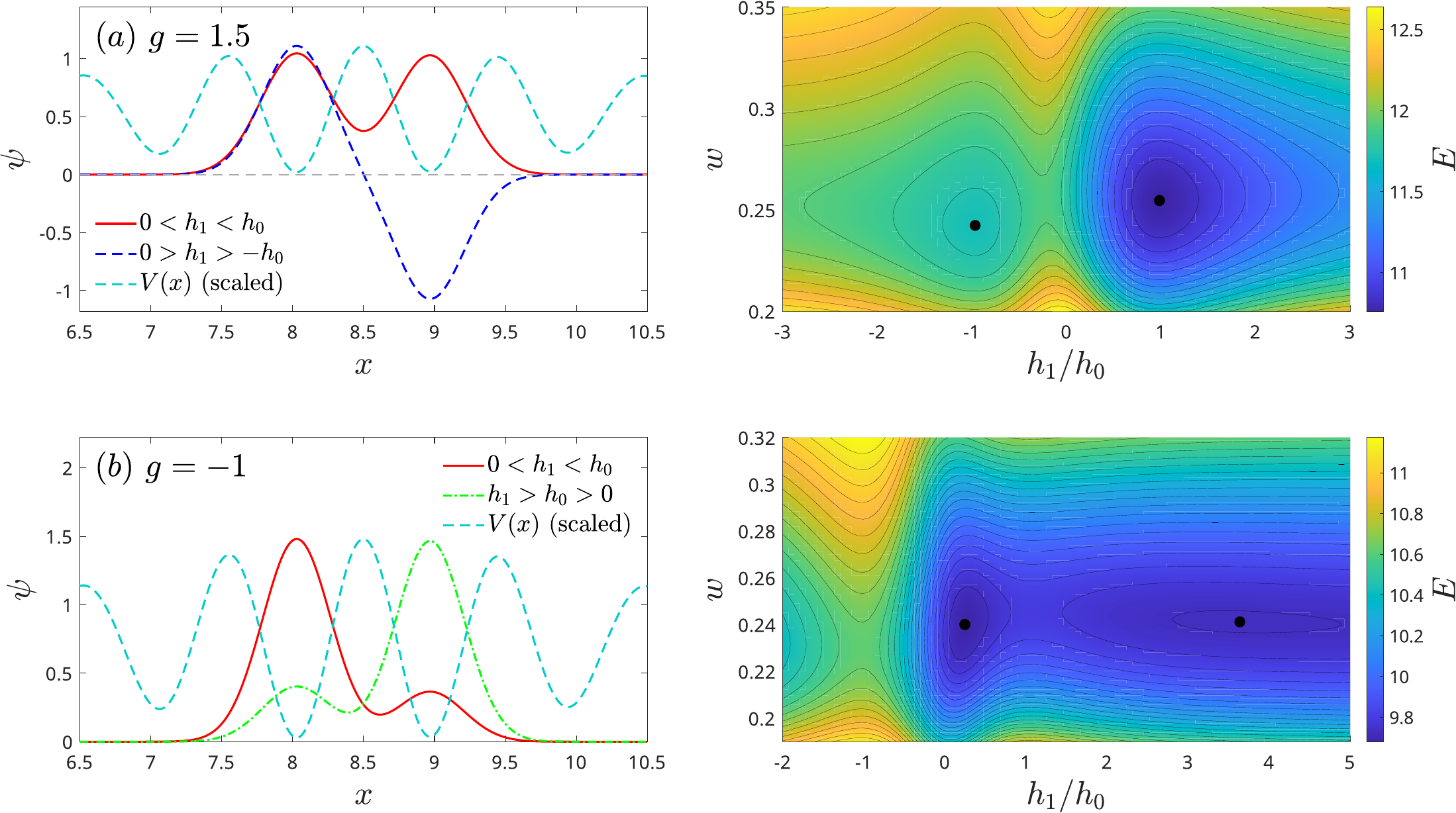}
\caption{Variational results illustrating the wave function profiles (left)
and color-coded energy $E$ (right) as functions of the variational
parameters $w$ and $b = h_{1}/h_{0}$, for $\protect\alpha = 2$. Panels (a)
and (b) correspond to self-repulsive ($g = 1.5$) and attractive ($g = -1$)
nonlinearities, respectively. Local energy minima are indicated by black
dots. The red solid line represents in-phase solutions with $0 < h_{1}/h_{0}
< 1$, the blue dashed line denotes spatially antisymmetric (out-of-phase)
solutions with $h_{1}/h_{0} < 0$, and the green dash-dotted line indicates a
second solution with broken symmetry ($h_{1}/h_{0} > 1$).}
\label{fig:E_w_b}
\end{figure*}

To identify local energy minima that can produce the VA solutions, we plot
the dependence of the energy on the variational parameters $w$ (width) and $%
b=h_{1}/h_{0}$ [relative amplitude, see Eq. (\ref{Gauss})] in Fig. \ref%
{fig:E_w_b}. It is seen that for large enough values of $|g|$, there are
two local minima instead of one. There exist critical values $g_{+}>0$ and $%
g_{-}<0$, which depend on the LI $\alpha $, such that for $g>g_{+}$ we have
out-of-phase solutions, with $h_{1}/h_{0}<0$ (in addition to regular ones
with $h_{1}/h_{0}>0$), as seen in panel (a) of Fig. \ref{fig:E_w_b}, and for
$g<g_{-}$, panel (b) also exhibits two solutions, with \textit{broken
symmetry} between the two potential wells ($h_{1}/h_{0}<1$ and $%
h_{1}/h_{0}>1 $). Spontaneous symmetry breaking is a well-known property of
NLSE solutions with two-well potentials and self-attractive nonlinearity
\cite{SSB1,SSB2}.

It should be noted that, unlike the case of the truly symmetric double-well
potential \cite{Malomed2016}, states with broken symmetry below the critical
value $g_-$ do not emerge from the single point at $h_1/h_0 = 1$, and the
product of their relative amplitudes is less than 1. Additionally, the
absolute value of the relative amplitude $|h_1/h_0|$ of the solutions that
exist for $g > g_-$ (in particular, out-of-phase states, which appear above $%
g_+$) is less than 1, in contrast to the case of the symmetric potential,
where this value is exactly 1. \textcolor{black}{ More detailed results for
states in the symmetric potential are presented in the Supplemental Material
of this article. As anticipated, the results for the periodic potential
closely resemble the characteristics of in-phase, out-of-phase, and
asymmetric solutions previously analyzed in the symmetric double-well
potential \cite{Malomed2008,Malomed2016}.}

\begin{figure*}[tbp]
\includegraphics[width=\textwidth]{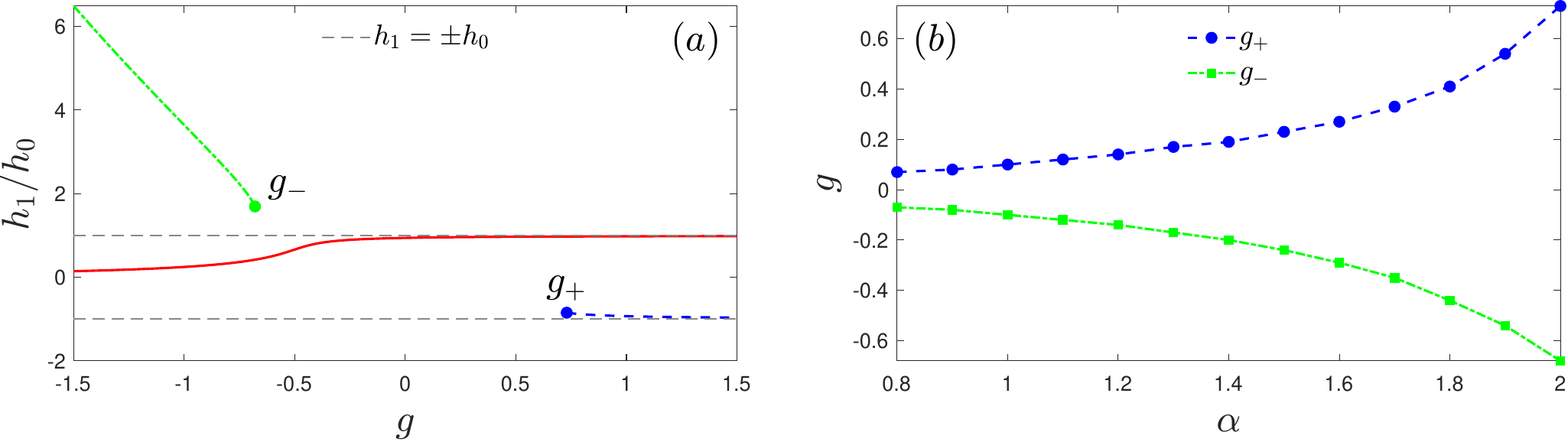}
\caption{(a) The variational parameter, the peak-amplitude ratio $%
b=h_{1}/h_{0}$, as a function of the coupling constant $g$, in the case
of the nonfractional diffraction ($\protect\alpha =2$) and normalization $%
N=1$. (b) The dependence of the critical values of the coupling constant, $%
g_{+}$ and $g_{-}$, on LI $\protect\alpha $. The second (spatially
antisymmetric, out-of-phase) solution, with $h_{1}/h_{0}<0$ (blue dashed
lines in both panels), appears at $g>g_{+}$, and the second solution with
broken symmetry ($h_{1}/h_{0}>1$, green dash-dotted lines in both panels)
appears at $g<g_{-}$. The red solid line indicates the in-phase solutions
with $0<h_{1}/h_{0}<1$.}
\label{fig:b_g}
\end{figure*}

{\color{black}
The properties of the VA-predicted two-peak states are further detailed in Fig. \ref{fig:b_g}. Panel (a) depicts the relative peak heights of the wave function in the bound states as a function of $g$ while panel (b) presents the critical values $g_{\pm}$ as functions of the LI parameter $\alpha$. The critical values $g_{\pm}$ approach zero as $\alpha$ decreases, reflecting the fact that weaker nonlinearity is sufficient to induce spontaneous symmetry breaking when competing with reduced diffraction (smaller $\alpha$). Notably, the values of $g_{+}$ and $\left|g_{-}\right|$ remain remarkably close across all considered values of $\alpha$. The criticality discussed here leads to the bifurcation of solutions, altering both the number and symmetry of localized states as $g$ crosses the critical values $g_{+}$ and $g_{-}$.
 As previously noted, no solutions exist for $\alpha < 1$ and sufficiently strong attractive nonlinearity ($g < 0$). Consequently, for $\alpha < 1$, the red and green lines in Fig. \ref{fig:b_g}(a) terminate at some critical value, $g_c(\alpha) < 0$. A comparison between Figs. \ref{fig:chi_plot_2D} and \ref{fig:b_g}(b) reveals that $g_-(\alpha) > g_c(\alpha)$, indicating that the point $(\alpha, g_-(\alpha))$ lies within the soliton existence region. Thus, solutions with broken symmetry exist even for small values of $\alpha$.}

 {\color{black}
For configurations with more than two peaks, it becomes impractical to construct 2D representations, such as the one in 
Fig. \ref{fig:E_w_b}, that effectively capture the emergence of secondary minima in the variational energy. However,
employing the nearest-neighbor approximation, the interaction between each pair of peaks can be analyzed independently, 
effectively reducing the multipeak problem to a series of two-peak cases. 
Naturally, the critical values of the nonlinearity differ for each pair of the density peaks.}

\subsection{Numerical solutions}

%%%%%%%----------------

To find stationary solutions numerically, we used a dissipative version of
FNLSE (\ref{eq:GPE}), including an artificial damping parameter $0<\gamma
\ll 1$:
\begin{equation}
(i-\gamma )\,\frac{\partial \widetilde{\Psi }}{\partial t}=\frac{1}{2}\left(
-\frac{\partial ^{2}}{\partial x^{2}}\right) ^{\alpha /2}\widetilde{\Psi }%
+V(x)\widetilde{\Psi }+\widetilde{g}|\widetilde{\Psi }|^{2}\widetilde{\Psi }%
-\mu \widetilde{\Psi }.  \label{eq:dampedGPE}
\end{equation}%
We use the damped version of GPE {\color{black}(DGPE)} to find stationary states rather than to
describe some physical dissipative process \cite{DGPE}.
{\color{black}For our simulations we used reasonably small value, $\gamma=0.05$, which gives fast convergence to the stationary state.}
%{\LARGE [It is relevant to add here a reference to this method.] }
Thus, simulating Eq. (\ref{eq:dampedGPE}) with the VA-predicted initial
condition, it is possible to achieve a numerically accurate stationary state
corresponding to chemical potential $\mu $.
{\color{black}Values of the chemical potential are obtained from
preliminary variational estimates to ensure that the numerical method
converges to a physically meaningful stationary state, since
choosing an arbitrary $\mu$ could lead to convergence towards a
state that does not correspond to the minimum of the energy functional}.
Although the norm of the wave
function and energy are not conserved in the course of the evolution
governed by Eq. (\ref{eq:dampedGPE}), the wave function eventually converges
to a stationary state corresponding to the chemical potential $\mu $. The
simulations pulled up when the integral residual, defined as
\begin{equation}
\delta \Psi =\int\limits_{-\infty }^{+\infty }\left\vert \left[
\frac12\left( -\frac{\partial ^{2}}{\partial x^{2}}\right) ^{\alpha
/2}+V+g|\Psi |^{2}-\mu \right] \Psi \right\vert ^{2}dx,
\end{equation}%
attained its minimum at some point in time, $t$ $=$ $t_{0}$.
{\color{black}A small residual indicates that the numerical solution satisfies the stationary equation \eqref{psi} with good accuracy,
meaning that the wave function, obtained via DGPE, closely approximates a true stationary state.
The combined approach of VA and DGPE guarantees that the final state obtained through our numerical simulations corresponds to an energy-minimizing configuration.}
Having obtained
the stationary state, we renormalized the wave function and the coupling
constant as follows:
\begin{equation}
\psi(x) =\widetilde{\Psi }(x,t_{0})/\sqrt{\widetilde{N}},~g=\widetilde{g}%
\widetilde{N},
\end{equation}
where $\widetilde{N}$ is the norm of the wave function at $t=t_{0}$, to
restore the adopted normalization, $N=1$.

{\color{black}
We have benchmarked the numerical results obtained via Eq. (\ref{eq:dampedGPE}) against two independent numerical methods: (i) the imaginary-time propagation (ITP) method \cite{ITP1,ITP2} and (ii) the modified squared-operator method (MSOM) \cite{yang2010nonlinear}. For small $g>0$, the ITP method yields solutions consistent with our approach. However, as $g$ increases, ITP convergence deteriorates due to the wave function spreading over a large spatial domain. Comparisons with MSOM reveal even higher accuracy at low nonlinearity, yet MSOM fails to converge reliably at stronger nonlinearities. Given its robust performance across a wide parameter range, we primarily adopt the approach based on Eq. (\ref{eq:dampedGPE}).}

\section{Dynamics of solitons}

\label{sec:dynamics} %%%%%%%----------------

\begin{figure}[h]
\includegraphics[width=0.48\textwidth]{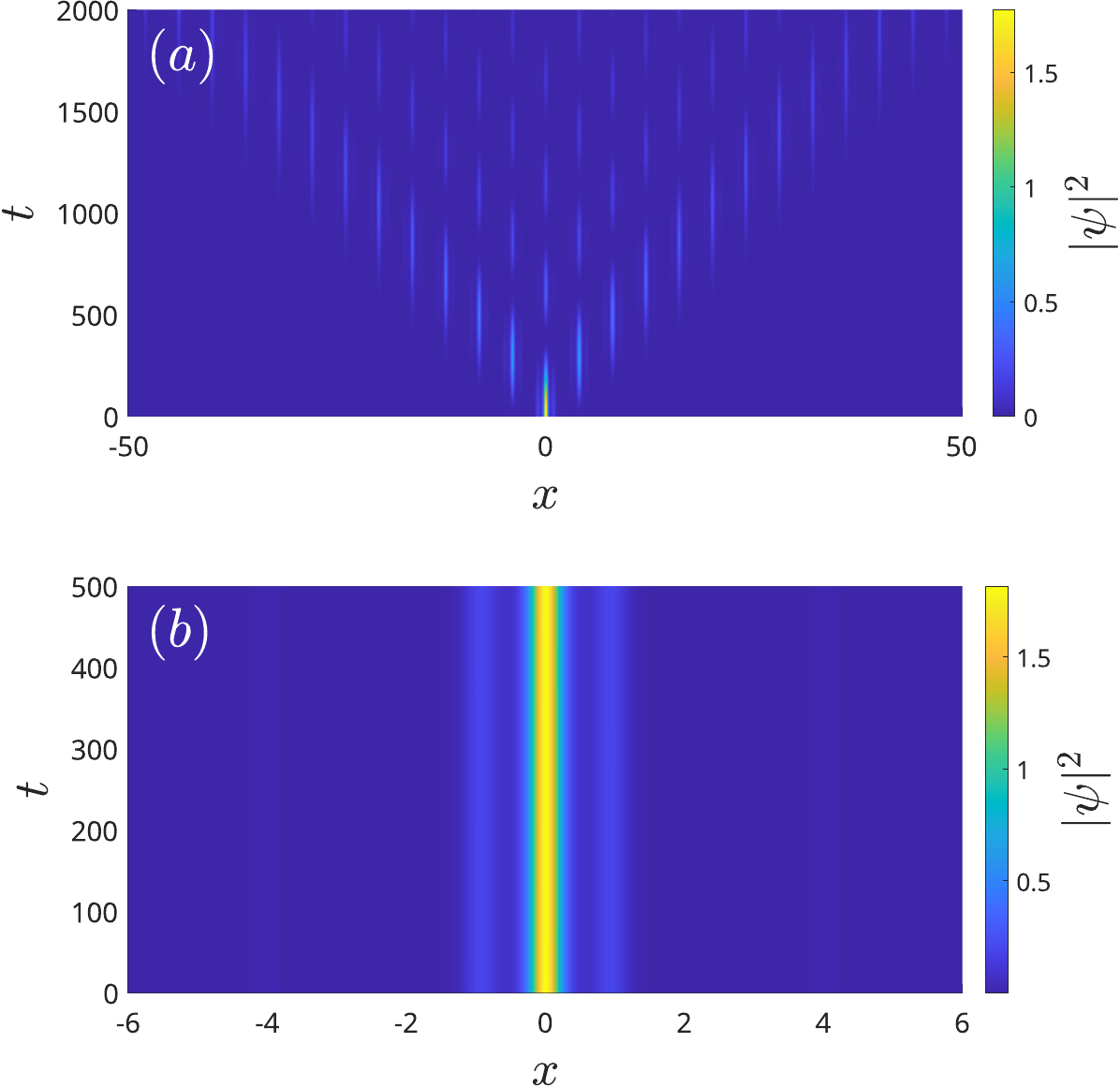}
\caption{The evolution of three-peak numerical solutions in the linear model
with normal diffraction ($g=0,\protect\alpha =2$) under the action of
the periodic (a) and quasiperiodic (b) OL potentials.}
\label{fig:dynamics_num_0}
\end{figure}

\textcolor{black}{The robustness of the bound states was examined by numerical simulations of their
evolution.
It is important to note that VA provides a good approximation for stationary states but does not guarantee their dynamical stability{\color{black},
which is not an inherent property of a stationary solution but depends
on its response to perturbations}.}
 First, Fig. \ref{fig:dynamics_num_0} shows the obvious difference
in the evolution of the numerical solutions in the linear model [$g=0$ in
Eq. (\ref{eq:GPE})] with the periodic and quasiperiodic OL potentials, in
the case of nonfractional diffraction $\alpha =2$. In accordance with the
commonly known principles, the wave function spreads out in the former case
and remains confined due to the AL effect in the latter situation.

% \begin{figure}[h]
%     \includegraphics[width=0.48\textwidth]{psi_t, itp, var_peaks=3, xmax=6, alpha=2.00, g=0.00, l1=2.000, l2=1.618, s1=2.00, s2=0.40, x0=0.000, duration=500}
%     \caption{Dynamics of the numerical solution for a non-interacting condensate with normal diffraction ($g=0,\alpha=2$).}
%     \label{fig:dynamics_num_1}
% \end{figure}

% To make sure that localization occurs not due to the depth of the potential, we verify that in the case of the rational relation between the wavelengths $\lambda_2/\lambda_1$, condensate will not remain localized over time (see Fig. \ref{fig:dynamics_num_0}).

\begin{figure}[h]
\includegraphics[width=0.48\textwidth]{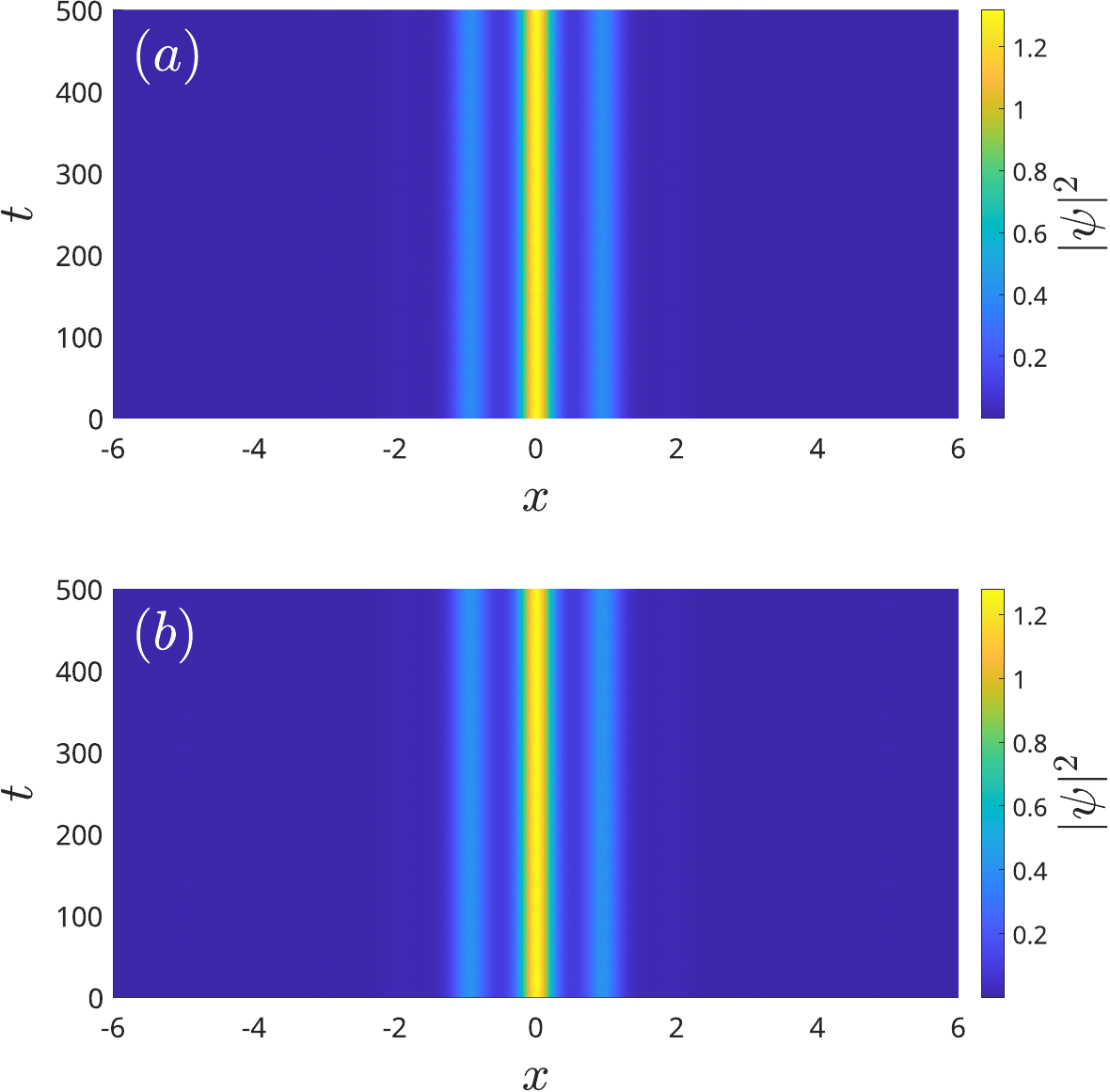}
\caption{The evolution of the three-peak numerical solutions in the case of
the self-repulsion and normal (nonfractional) diffraction ($g=1,\protect%
\alpha =2$), with norm $N=1$, under the action of the periodic (a) and
quasiperiodic (b) OL potentials.}
\label{fig:dynamics_num_1}
\end{figure}

On the other hand, it is shown in Fig. \ref{fig:dynamics_num_1} that, under
the action of the repulsive self-interaction and nonfractional diffraction (%
$\alpha =2$), not only the quasiperiodic potential but also the periodic one
make it possible to create stable localized states, actually, as the gap
solitons.

\begin{figure}[h]
\includegraphics[width=0.48\textwidth]{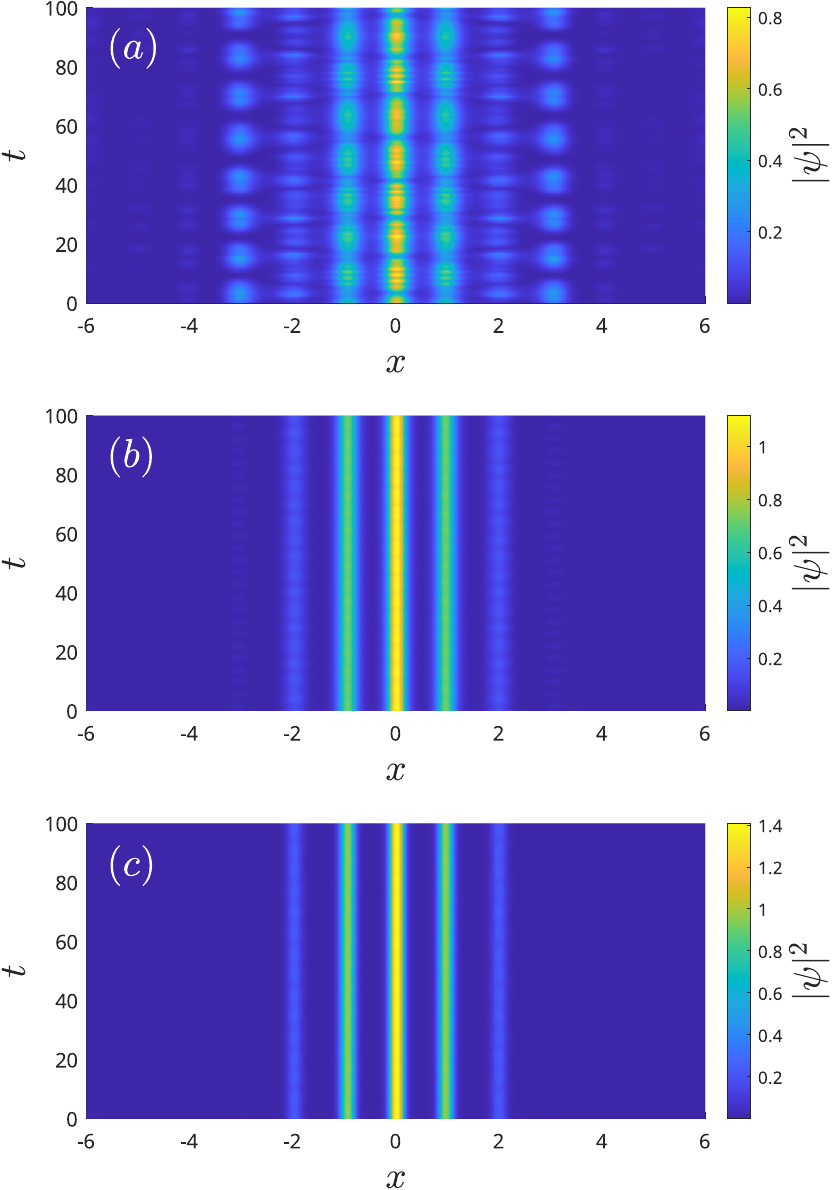}
\caption{The evolution of the five-peak numerical solutions in the case of strong self-repulsion ($g=5$) and different values of LI $\protect\alpha
$, with norm $N=1$: (a) $\protect\alpha=2$, (b) $\protect\alpha=1.5$, and (c) $%
\protect\alpha=1$.}
\label{fig:dynamics_num5_5}
\end{figure}

In the case of the fractional diffraction ($\alpha <2$), with the same
strength of the self-repulsion (coupling constant), $g=1$, the bound states
demonstrate similar robustness. Next, we address FNLSE with stronger
repulsive interaction ($g=5$), as in this case Fig. \ref{fig:dynamics_num5_5}
exhibits a {difference in dynamics} between different values of LI $%
\alpha $. Furthermore, we here produce the results for five-peak bound
states to explore the robustness of more complex bound states. It is seen
that the bound states are notably {more robust} for the fractional
diffraction, with $\alpha <2$.

\textcolor{black}{As shown in Figs. \ref{fig:dynamics_num_0} and \ref%
{fig:dynamics_num_1}, the distinction between periodic and quasiperiodic potentials becomes evident for very small nonlinearity. When the local nonlinearity is sufficiently strong, it dominates over the long-range effects introduced by the periodicity or quasiperiodicity of the lattice. Consequently at large nonlinearities, the localization 
exhibits minimal differences between the periodic and quasiperiodic potential cases.}

\begin{figure}[h]
\includegraphics[width=0.48\textwidth]{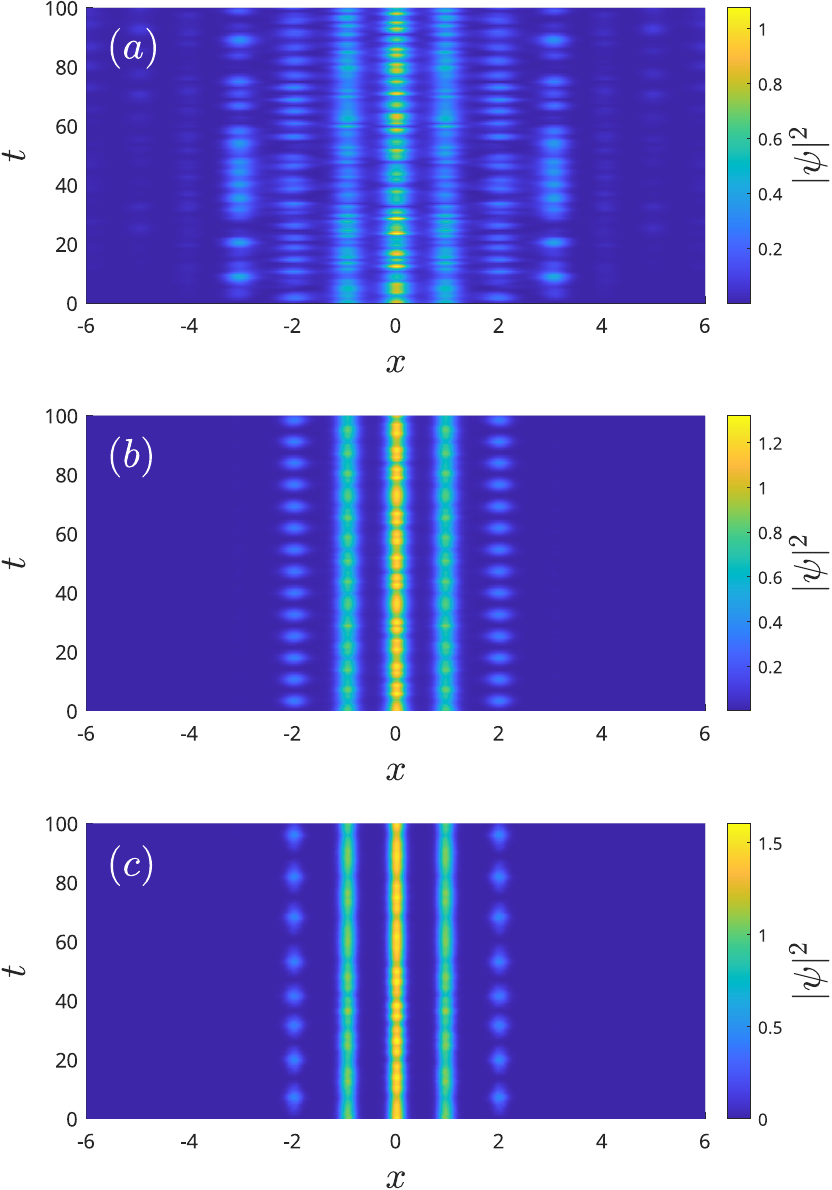}
\caption{The evolution of the three-peak VA-predicted solutions under the
action of strong self-repulsion ($g=5$) and different values of LI $\protect%
\alpha $, with norm $N=1$: (a) $\protect\alpha=2$, (b) $\protect\alpha=1.5$,
and (c) $\protect\alpha=1$.}
\label{fig:dynamics_var3_5}
\end{figure}

While the results displayed in Figs. \ref{fig:dynamics_num_0}, \ref%
{fig:dynamics_num_1}, and \ref{fig:dynamics_num5_5} were obtained in the
fully numerical form, it is also relevant to check the evolution of the
VA-predicted three-peak modes. It is seen in Fig. \ref{fig:dynamics_var3_5}
that, under the action of the fractional diffraction, with $\alpha =1.5$
and, especially, $\alpha =1$, the ensuing dynamics is essentially more
steady than in the case of the nonfractional diffraction, $\alpha =2$,
i.e., the VA accuracy improves with the decrease of $\alpha $.

\textcolor{black}{The VA effectively identifies stationary solutions by minimizing the energy functional. However, their stability depends on nonlinear dynamics, which may amplify perturbations. Using variational solutions as initial conditions, we probed their dynamical stability. It is crucial to complement the VA with direct numerical simulations to gain a comprehensive understanding of the intricate interplay between nonlinearity and dynamical stability. {\color{black}
It should be emphasized that, in our case, no external perturbations were introduced, apart from those arising from the inherent inaccuracy of the variational solutions. While numerical precision is necessarily finite, its effect on the system's evolution remains negligible.}}

%{\color{red}It is worth noting that observed instability of certain solutions in time evolution is not solely due to the inaccuracy of the VA - it can also result from the intrinsic instability of the exact stationary solution itself.}}
%%%%%%%%%%%%%%%%%%%%%%%%%%%%%%%%%%%%%%

\section{Conclusion}

\label{sec:conclusions} %%%%%%%%%%%%%%%%%%%%%%%%%%%%%%%%%%%%%%
We have investigated the interplay between the fractional diffraction and 1D
quasiperiodic potentials in shaping \textcolor{black}{the general properties
of soliton structures and their} dynamics. Using the VA (variational
approximation) and numerical methods, we have produced self-trapped states
and identified their stability, varying the LI (L\'{e}vy index), which
determines the fractional diffraction. These findings not only advance the
understanding of the nonlinear wave propagation in fractional and disordered
media but also expand potential applications of solitons in optical
data-processing schemes by revealing conditions for {improving the stability
of the solitons.}

It is relevant to extend the analysis for the compound states including two
or more of the localized modes considered above. Promising possibilities are to
consider the interplay of the fractional diffraction with other forms of
nonlinearity (in particular, nonpolynomial terms \cite{Luca,Delgado}, which
are produced by strong confinement applied to BEC in the transverse
directions) and, eventually, higher-dimensional settings \cite{book}. In
particular, it may be interesting to construct two- and three-dimensional
localized modes with embedded vorticity. {\color{black} Moreover, aperiodic two-dimensional lattice potentials,
cognate to quasiperiodic ones, such as the Sierpi\'{n}ski gasket, are used as the
basis for the creation of topological photonic setups such as optical topological isolators (TIs)
and higher-order TIs \cite{TI1,TI2}}.

\section*{Acknowledgments}

The work of B.A.M. was supported, in part, by the Israel Science Foundation
through grant No. 1695/22.
L.S. and A.I.Y. thank also the following Italian MUR projects for
financial support: “PRIN 2022: Quantum Atomic Mixtures:
Droplets, Topological Structures, and Vortices” and “Dipartimento di Eccellenza DFA: Quantum Frontiers”.

\section*{Appendix: Generalized variational analysis}\label{sec:general_variational_method}

Here we present calculations of the integrals with the multipeak ansatz:
\begin{equation}
    \psi=\sum_j h_j\exp\left(-\frac{(x-x_j)^2}{2w^2}\right).
\end{equation}
The corresponding energy functional is
\begin{gather}
    E=\int\limits_{-\infty}^{+\infty} dx\left[\psi^*\,\frac{1}{2}\left(-\frac{\partial^2}{\partial x^2}\right)^{\alpha/2}\psi+V(x)|\psi|^2+\frac{g}{2}\,|\psi|^4\right]\notag\\
    =\frac{1}{2\pi}\int\limits_{0}^{+\infty}p^{\alpha}|\mathcal{F}\psi|^2dp+\sum_{i=1}^{2}A_i\int\limits_{-\infty}^{+\infty}\sin^2\left(\kappa_ix\right)|\psi|^2dx\notag\\
    +\frac{g}{2}\int\limits_{-\infty}^{+\infty}|\psi|^4dx,
\end{gather}
where $\mathcal{F}$ is the symbol of the Fourier transform.
We consider only the overlapping between neighboring peaks, denoting $b_j=h_j/h_0$,
where $h_0$ is the amplitude of one of the peaks (for instance, the central one).
We have variational parameters $w$ and $\{b_{j\neq 0}\}$, whose total number is
equal to the number of peaks. For the norm of the ansatz we obtain
\begin{gather}
    N=\int\limits_{-\infty}^{+\infty}|\psi|^2dx\notag\\
    =\sum_{j,k}h_jh_k\int\limits_{-\infty}^{+\infty}\exp\left(-\frac{(x-x_j)^2+(x-x_k)^2}{2w^2}\right)dx\notag\\
    =\sum_{j,k}\left\{h_jh_k\exp\left(-\frac{(x_j-x_k)^2}{4w^2}\right)\right.\notag\\
    \left.\times\int\limits_{-\infty}^{+\infty}\exp\left[-\frac{1}{w^2}\left(x-\frac{x_j+x_k}{2}\right)^2\right]dx\right\}\notag\\
    =\sqrt{\pi}w\sum_{j,k}h_jh_k\varepsilon_{jk}(w)=\sqrt{\pi}wh_0^2f_N,
\end{gather}
where
\begin{gather}
    f_N=1+\sum_{j\neq 0}b_j^2+2\sum_{\langle j,k\rangle}b_jb_{k}\varepsilon_{jk}(w),\\
    \varepsilon_{jk}(w)=\exp\left(-\frac{(x_j-x_k)^2}{4w^2}\right),
\end{gather}
and $\displaystyle\sum_{\langle j,k\rangle}$ denotes the sum over the neighboring
indices. By fixing the norm $N$, we can eliminate $h_0$:
\begin{equation}
    h_0^2=\frac{N}{\sqrt{\pi}wf_N}.
\end{equation}
The corresponding kinetic-energy term is
\begin{gather}
    \frac{1}{2\pi}\int\limits_{0}^{+\infty}p^{\alpha}|\mathcal{F}\psi|^2dp\notag\\
    =\frac{1}{2\pi}\int\limits_{0}^{+\infty}p^{\alpha}\left|\sum_{j}h_j\int\limits_{-\infty}^{+\infty}\exp\left(-\frac{(x-x_j)^2}{2w^2}\right)e^{-ipx}dx\right|^2dp\notag\\
    =\frac{1}{2\pi}\int\limits_{0}^{+\infty}p^{\alpha}\left|\sum_{j}h_je^{-ipx_j}\sqrt{2\pi}we^{-p^2w^2/2}\right|^2dp\notag\\
    =w^2\sum_{j,k}h_jh_k\int\limits_{0}^{+\infty}p^{\alpha}e^{-p^2w^2}\cos\Big(p(x_j-x_k)\Big)dp\notag\\
    =\frac{\Gamma\left(\frac{\alpha+1}{2}\right)}{2w^{\alpha-1}}\sum_{j,k}h_jh_k\chi_{jk}^{(\alpha)}(w)
    =\frac{\Gamma\left(\frac{\alpha+1}{2}\right)N}{2\sqrt{\pi}w^{\alpha}}\,\frac{f_K}{f_N},
\end{gather}
where
\begin{gather}
    f_K=1+\sum_{j\neq 0}b_j^2+2\sum_{\langle j,k\rangle}b_jb_{k}\chi_{jk}^{(\alpha)}(w),\\
    \chi_{jk}^{(\alpha)}(w)=\,_1F_1\left(\frac{\alpha+1}{2},\frac{1}{2},-\frac{(x_j-x_k)^2}{4w^2}\right).
\end{gather}
The potential-energy term is
\begin{gather}
    \sum_{i=1}^{2}A_i\int\limits_{-\infty}^{+\infty}\sin^2\left(\kappa_ix\right)|\psi|^2dx\notag\\
    =\sum_{i=1}^{2}A_i\sum_{j,k}\left\{h_jh_k\varepsilon_{jk}(w)\int\limits_{-\infty}^{+\infty}\sin^2(\kappa_ix)\right.\notag\\
    \left.\times\exp\left[-\frac{1}{w^2}\left(x-\frac{x_j+x_k}{2}\right)^2\right]dx\right\}\notag\\
    =\frac{\sqrt{\pi}w}{2}\sum_{i=1}^{2}A_i\sum_{j,k}h_jh_k\varepsilon_{jk}(w)\sigma_{jk}^{(i)}(w)\notag\\
    =\frac{N}{2f_N}\sum_{i=1}^{2}A_if_P^{(i)},
\end{gather}
where
\begin{gather}
    f_P^{(i)}=\sigma_{00}^{(i)}(w)+\sum_{j\neq 0}b_j^2\sigma_{jj}^{(i)}(w)\notag\\
    +2\sum_{\langle j,k\rangle}b_jb_k\varepsilon_{jk}(w)\sigma_{jk}^{(i)}(w),\\
    \sigma_{jk}^{(i)}(w)=1-e^{-\kappa_i^2w^2}\cos\Big(\kappa_i(x_j+x_k)\Big).
\end{gather}
The interaction-energy term is
\begin{gather}
    \frac{g}{2}\int\limits_{-\infty}^{+\infty}|\psi|^4dx=\frac{g}{2}\sum_{j_1,j_2,j_3,j_4}\left\{\left(\prod_{k=1}^{4}h_{j_k}\right)\right.\notag\\
    \left.\times\int\limits_{-\infty}^{+\infty}\exp\left(-\frac{1}{2w^2}\sum_{k=1}^{4}(x-x_{j_k})^2\right)dx\right\}\notag\\
    =\frac{\sqrt{\pi}gw}{2\sqrt{2}}\sum_{j_1,j_2,j_3,j_4}\Bigg\{\left(\prod_{k=1}^{4}h_{j_k}\right)\notag\\
    \left.\times\exp\left[-\frac{1}{2w^2}\left(\sum_{k=1}^{4}x_{j_k}^2-\frac{1}{4}\left(\sum_{k=1}^{4}x_{j_k}\right)^2\right)\right]\right\}\notag\\
    =\frac{gN^2}{2\sqrt{2\pi}w}\,\frac{f_I}{f_N^2},
\end{gather}
where
\begin{gather}
    f_I=1+\sum_{j\neq 0}b_j^4+4\sum_{\langle j,k\rangle}b_jb_k\left(b_j^2+b_k^2\right)\varepsilon_{jk}^{3/2}(w)\notag\\
    +6\sum_{\langle j,k\rangle}b_j^2b_k^2\varepsilon_{jk}^2(w).
\end{gather}
We thus obtain the expression for the total energy accounting on the interaction
between neighboring sites:
\begin{gather}
    E=\frac{\Gamma\left(\frac{\alpha+1}{2}\right)N}{2\sqrt{\pi}w^{\alpha}}\,\frac{f_K}{f_N}+\frac{N}{2f_N}\sum_{i=1}^{2}A_if_P^{(i)}+\frac{gN^2}{2\sqrt{2\pi}w}\,\frac{f_I}{f_N^2}\notag\\
    =\frac{N}{2f_N}\left[\frac{\Gamma\left(\frac{\alpha+1}{2}\right)}{\sqrt{\pi}w^{\alpha}}\,f_K+\sum_{i=1}^{2}A_if_P^{(i)}+\frac{gN}{\sqrt{2\pi}w}\,\frac{f_I}{f_N}\right].
    \label{eq:E_var_Npeaks}
\end{gather}

\begin{figure}[h]
\includegraphics[width=0.48\textwidth]{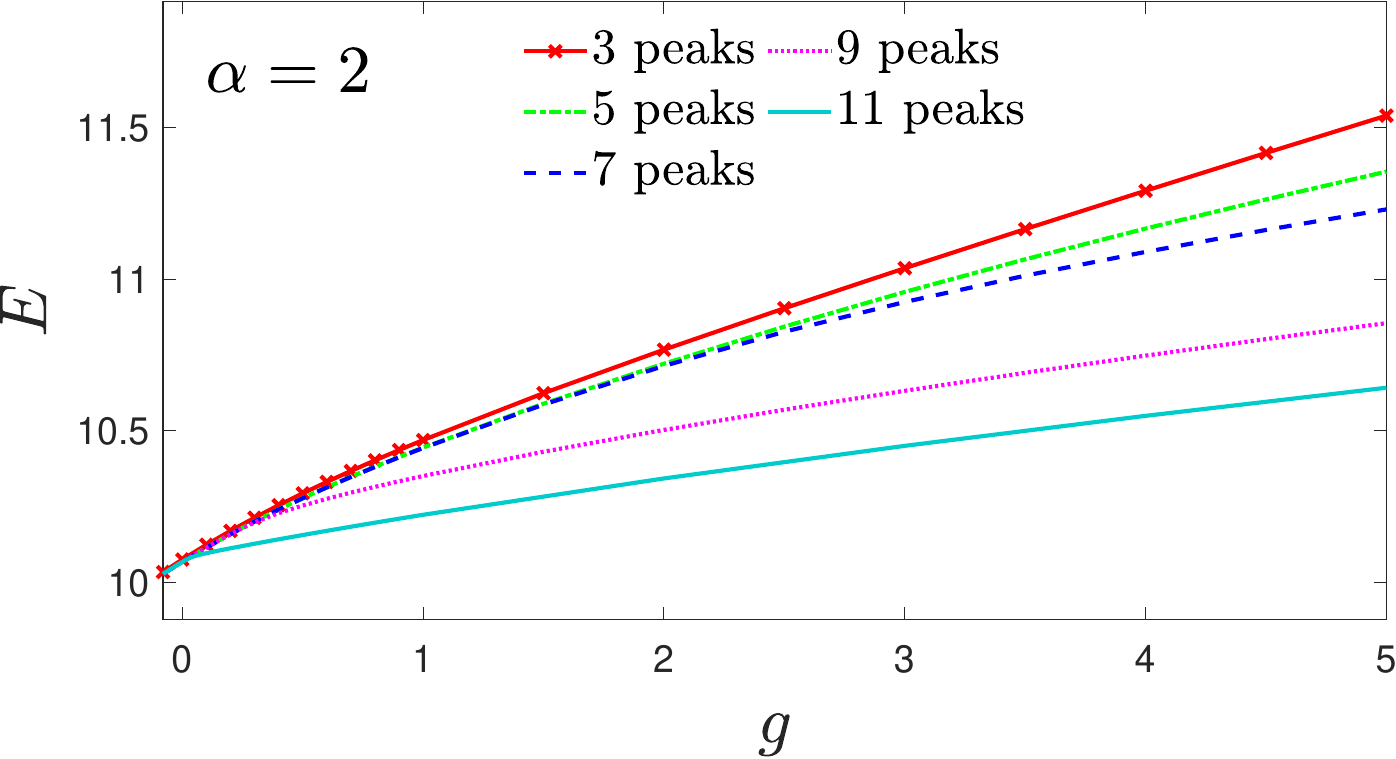}
\caption{VA results for the dependence of energy $E$ [see Eq. (\protect
\ref{eq:E_var_Npeaks})] on the coupling constant $g>0$ (self-repulsion) for
ansatze with different numbers of peaks [see Eq. (A1)] under normalization $N=1$ in the case of normal diffraction $(%
\protect\alpha =2)$.}
\label{fig:E_g}
\end{figure}

In Fig. \ref{fig:E_g} we see that an ansatz with a greater number
of peaks produces lower energy for a fixed value of the coupling constant $g$%
. This property reflects the fact that every additional peak increases the
number of degrees of freedom of the trial function and thus gives an option
to reduce the predicted value of the energy functional.

\begin{figure}[h]
\includegraphics[width=0.48\textwidth]{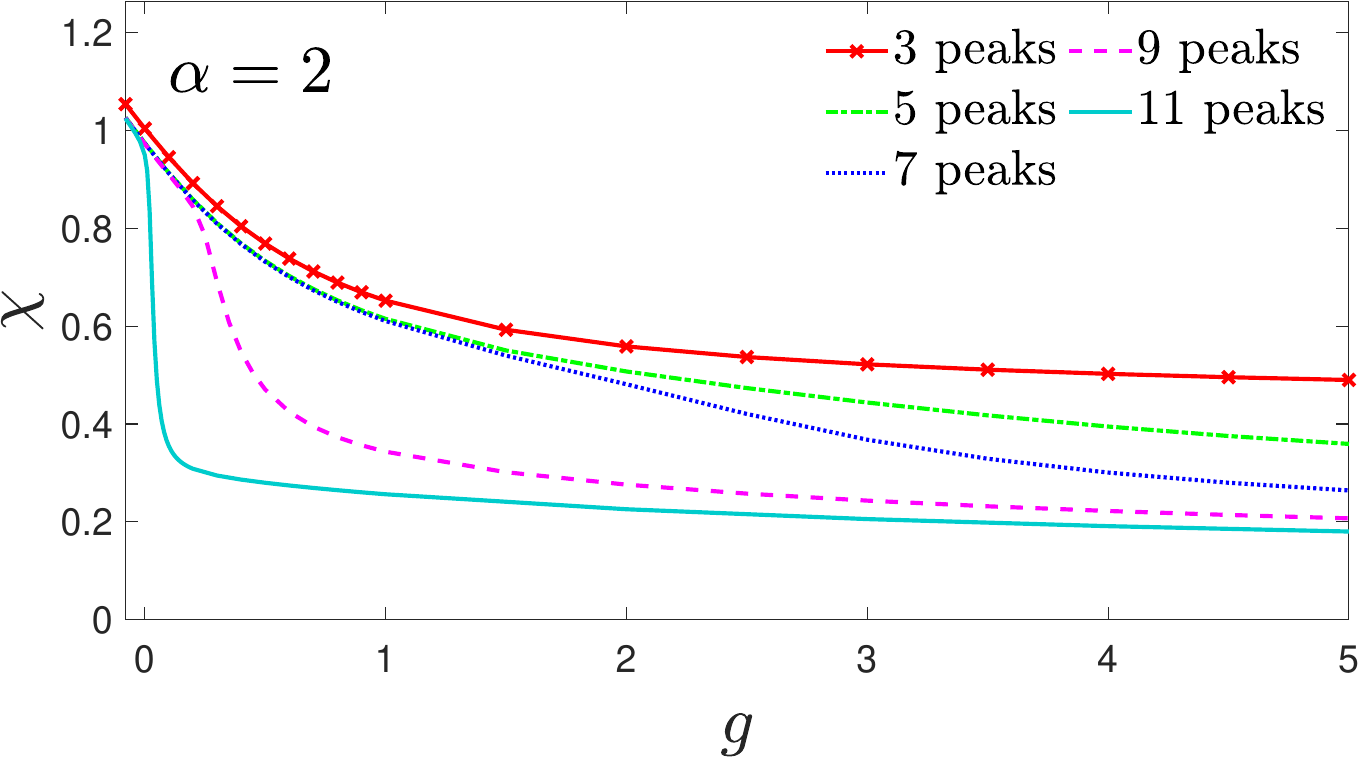}
\caption{VA results for ansatze with different numbers of
peaks, which produce the dependence of the form factor $\protect\chi $, see
Eq. \eqref{eq:chi}, on the coupling constant $g>0$ in the case of normal
diffraction $\protect\alpha =2$ and normalization $N=1$.}
\label{fig:chi_g}
\end{figure}

Figure \ref{fig:chi_g} shows that as the coupling constant $g>0$ increases (in
the case of the self-repulsion), the form factor $\chi $ decreases for each
fixed number of peaks of the ansatz, signifying the growth of the satellite
peaks.

%\section{Damped Gross-Pitaevskii equation}\label{sec:DGPE}

% {\LARGE [It is necessary to define exactly what is meant by $N$ here -- the final norm produced by the simulations of Eq. (\ref{eq:dampedGPE}).] }
% \end{appendix}
\section*{Supplemental Material}

\subsection{Two-hump states in periodic potential}

Recall the expression for the optical lattice potential:
\begin{equation}
    V(x)=\sum_{j=1}^{2}A_{j}\sin ^{2}\left( \frac{2\pi }{\lambda _{j}}\,x\right).
    \label{eq:V_sin}
\end{equation}

Here we present the results of the variational analysis (VA) with two-peak ansatz for the specific case of periodic potential with $A_2=0$ in Eq. (\ref{eq:V_sin}). As expected, the results for the periodic potential are very similar to the properties of the in-phase, out-of-phase symmetric and asymmetric solutions previously investigated in Refs.
\cite{Malomed2008,Malomed2016} for the symmetric double-well potential.
\begin{figure}[h]
\includegraphics[width=0.48\textwidth]{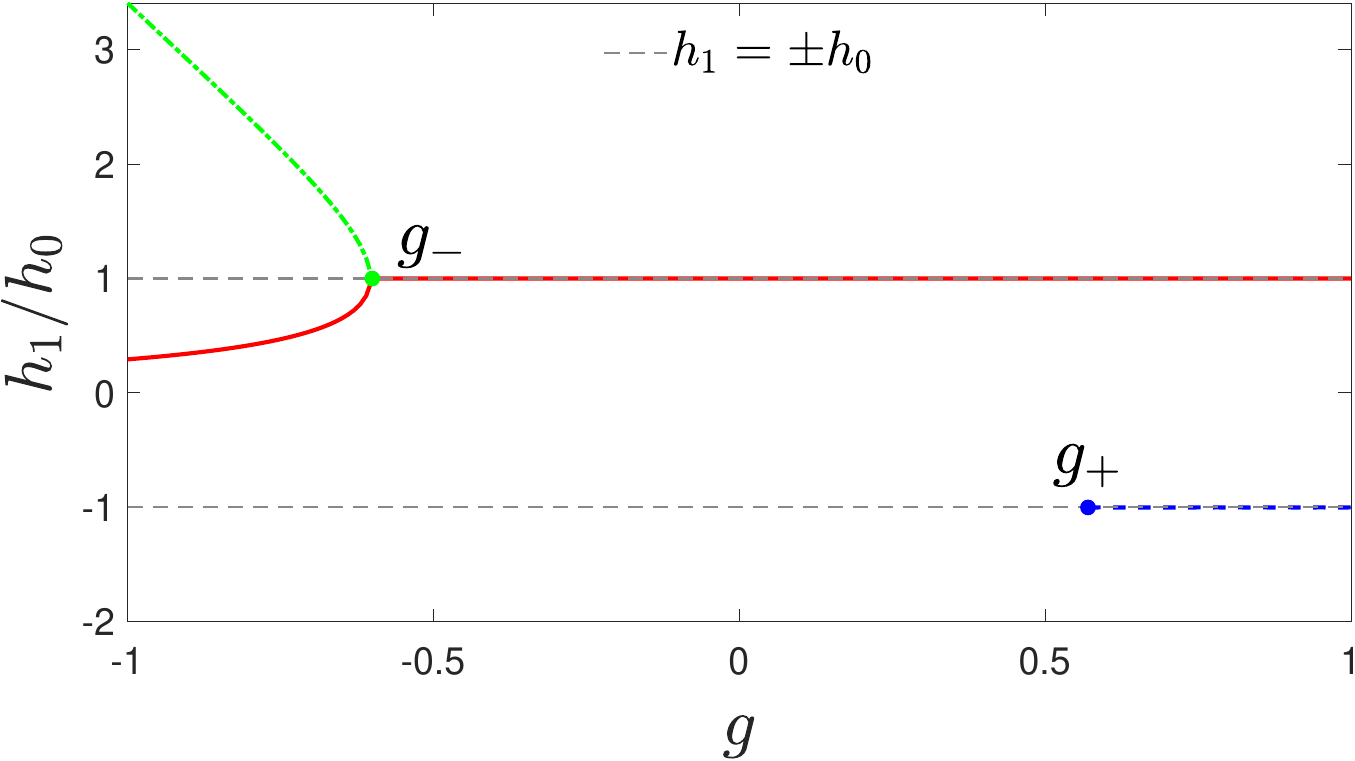}
\caption{The VA-predicted dependence of the peak-amplitude ratio ($%
b=h_{1}/h_{0}$) on coupling constant $g$ in the case of the non-fractional
diffraction ($\protect\alpha =2$) and normalization $N=1$ in the periodic potential. The second (spatially antisymmetric)
solution, with $h_{1}/h_{0}<0$ (blue dashed line), appears at $g>g_{+}$, and the second
solution with broken symmetry ($h_{1}/h_{0}>1$, green dash-dotted line) appears
at $g<g_{-}$.}
\label{fig:b_g_periodic}
\end{figure}

As shown in Fig. \ref{fig:b_g_periodic}, the symmetric in-phase state (red curve) with $h_1/h_0 = 1$ bifurcates into two asymmetric in-phase states (green and red curves) for $g < g_-$. Notably, the product of the ratios $(h_1/h_0)$ for these two branches equals unity, consistent with the preservation of mirror symmetry in the two-hump states of the potential.

\begin{figure*}[tbp]
\includegraphics[width=0.75\textwidth]{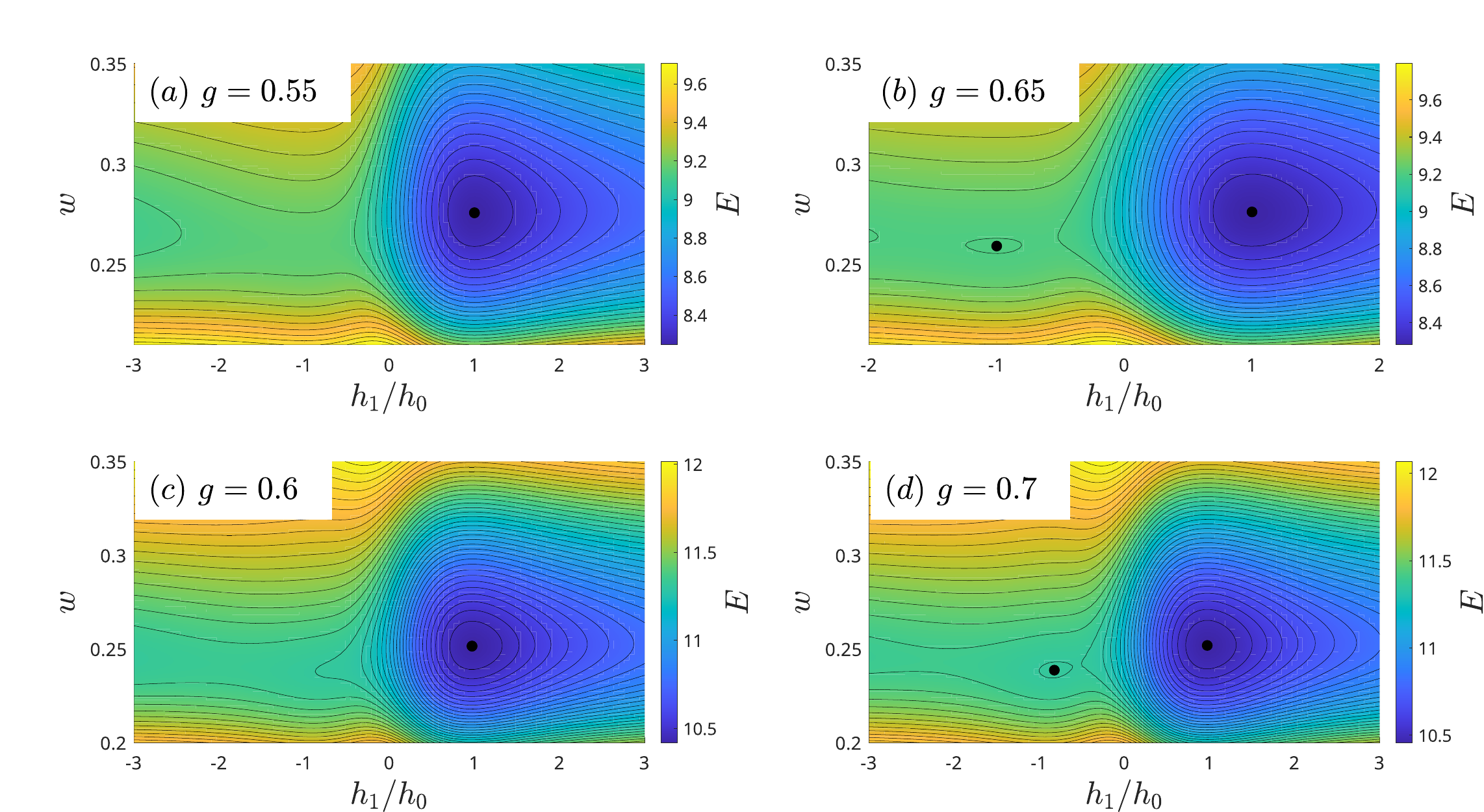}
\caption{The VA results illustrating the appearance of the second local minimum of energy $E$
in the case of the non-fractional diffraction ($\protect%
\alpha =2$) for the self-repulsive nonlinearity near the critical value $g_+$.
Local energy minima are marked by black dots. (a), (b) periodic potential; (c), (d) quasiperiodic potential.}
\label{fig:g+_rebuilding}
\end{figure*}

\begin{figure*}[tbp]
\includegraphics[width=0.75\textwidth]{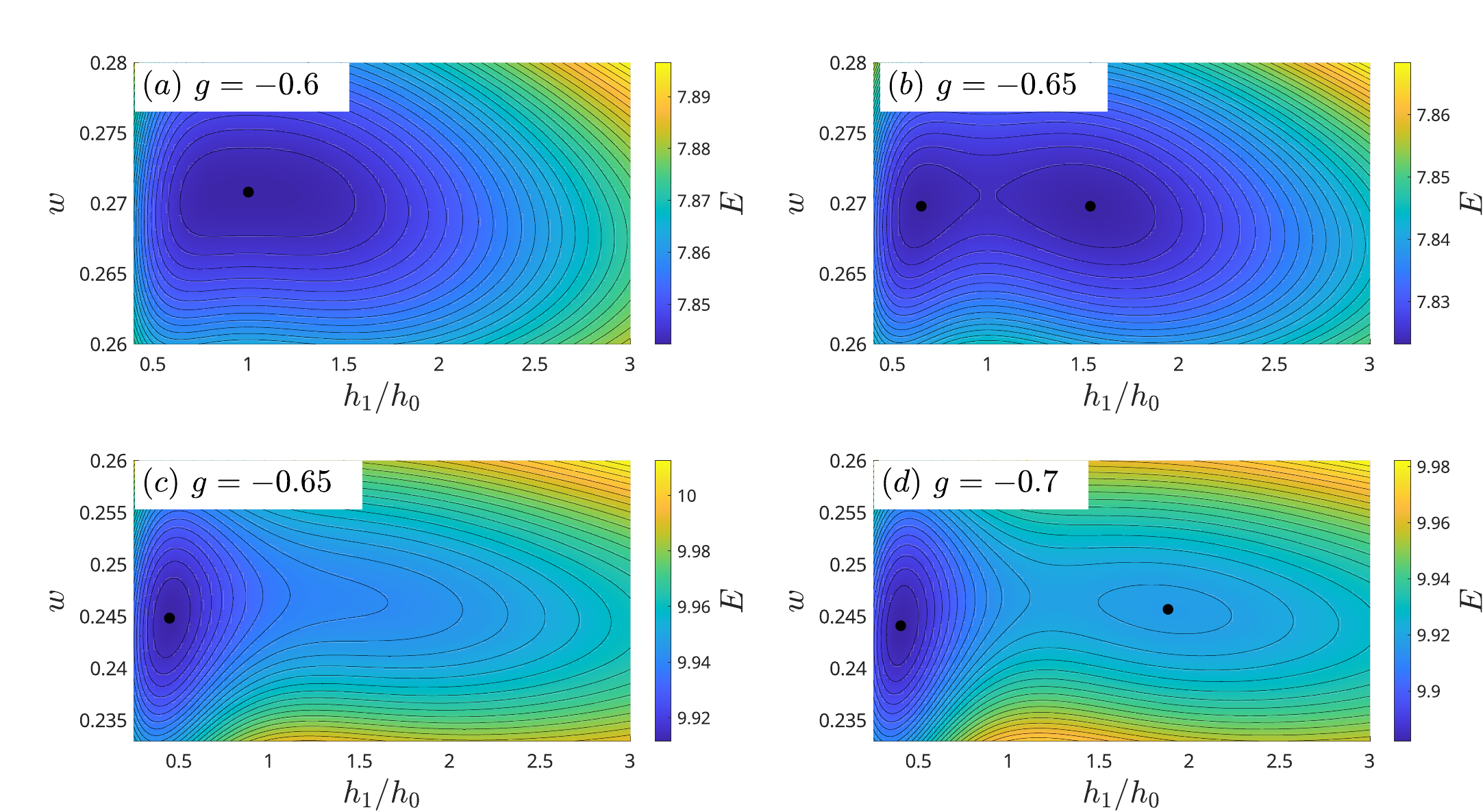}
\caption{The VA results illustrating the appearance of the second local minimum of energy $E$
in the case of the non-fractional diffraction ($\protect%
\alpha =2$) for the self-attractive nonlinearity near the critical value $g_-$.
Local energy minima are marked by black dots. (a), (b) periodic potential; (c), (d) quasiperiodic potential.}
\label{fig:g-_rebuilding}
\end{figure*}

The energy landscapes in the plane of the two variational parameters are depicted for different values of nonlinearity strength $g$, near the bifurcation points $g_+$ and $g_-$, in Fig. \ref{fig:g+_rebuilding} and Fig. \ref{fig:g-_rebuilding}, respectively.

Notably, the energy minima that correspond to the out-of-phase solutions are significantly less deep than the ones for the in-phase solutions, both in cases of periodic and quasiperiodic potentials (Fig. \ref{fig:g+_rebuilding}). Thus, from this energetic analysis follows that the out-of-phase solitons should be less robust than the in-phase states. These predictions are supported by our numerical simulation of the soliton dynamics. Also, the second solution with broken symmetry (with $h_1/h_0>1$) has greater energy than the first one (with $h_1/h_0<1$) in the case of the quasiperiodic lattice, in contrast to the periodic one, where respective energies are the same,
%Fig. \ref{fig:g-_rebuilding}
which reflects the fact that the second minimum of the quasiperiodic potential has a lower depth than the first one.

\subsection{Additional variational results}
\begin{figure}[h]
\includegraphics[width=0.48\textwidth]{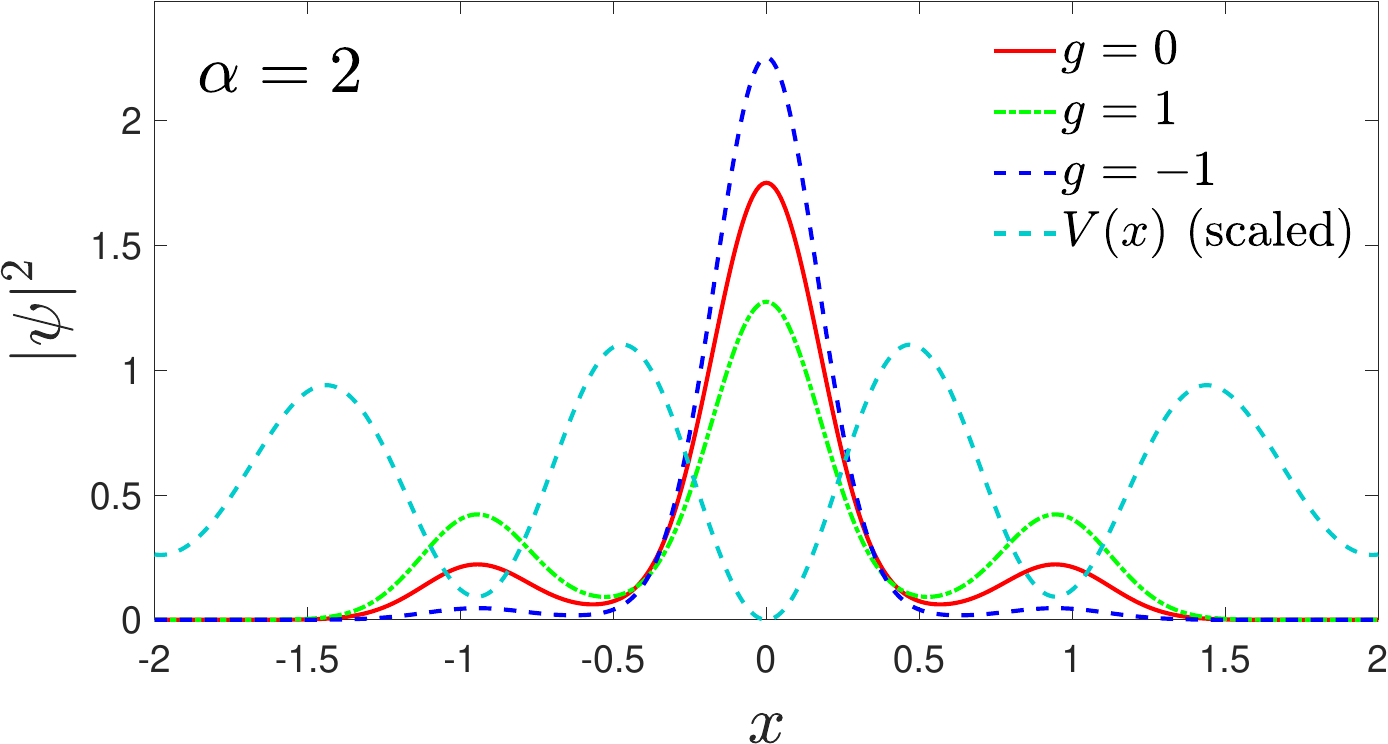}
\caption{Density profiles of the VA solutions for the three-peak ansatz for
different values of the normalized coupling constant ($g$), in the case of
the normal (non-fractional) diffraction, $\protect\alpha =2$.}
\label{fig:var_g}
\end{figure}

We now discuss the application of the variational method developed in this work to the well-studied case of a quasiperiodic lattice with non-fractional diffraction, $\alpha = 2$.
The density
profiles of the so-obtained VA solutions (bound states) are plotted, for
different values of coupling constant (nonlinearity coefficient) $g$, in
Fig. \ref{fig:var_g}. As expected, the height of the side peaks increases
with the increase of $g$ when repulsive interaction essentially modifies the soliton shape. In the limit of the strong repulsion (large $g$), the bound
states are well approximated by the Thomas-Fermi approximation, which
neglects the diffraction term and, therefore, is not affected by the value
of the $\alpha $ (this limit case is a well-known one, therefore it is not
presented here in detail).

\begin{figure}[h]
\includegraphics[width=0.48\textwidth]{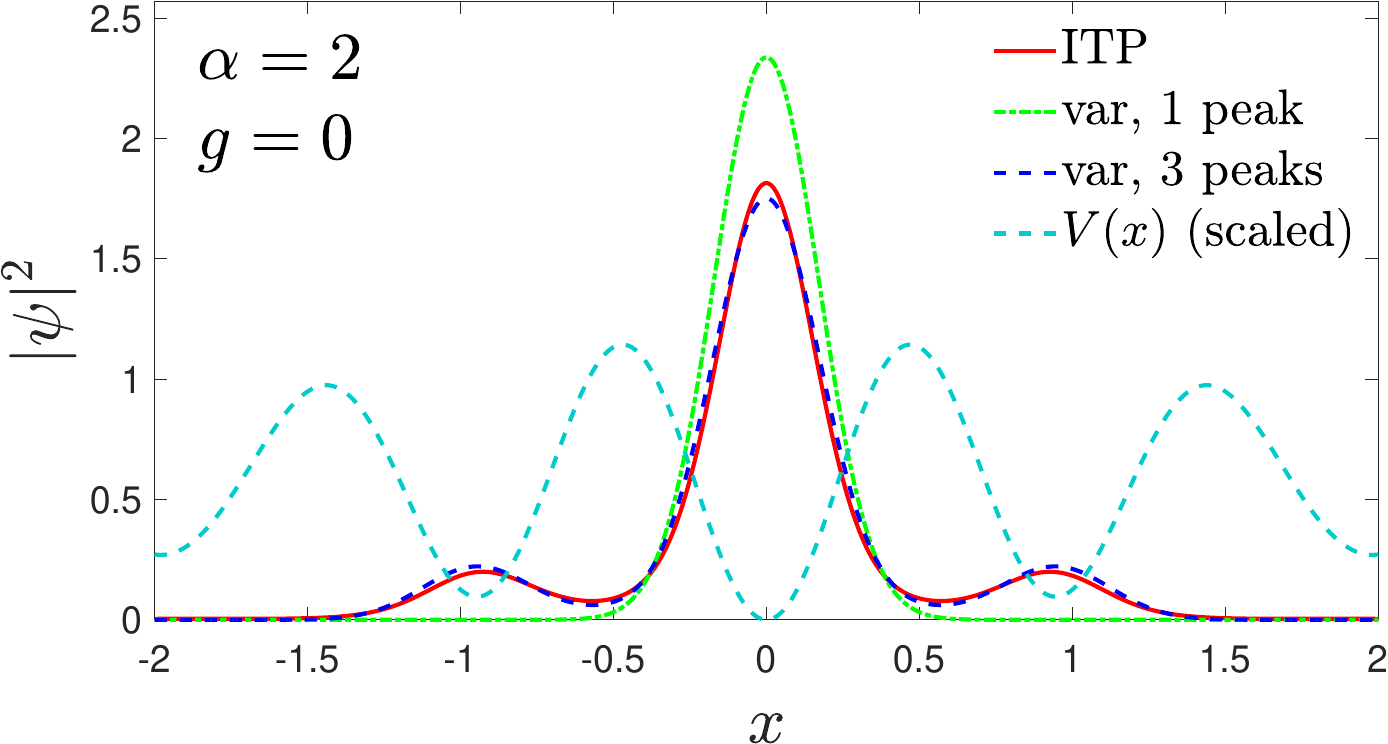}
\caption{VA solutions for single- and three-peak ansatz,
compared to the numerical solution, for the non-interacting condensate with
normal diffraction $(g=0,\protect\alpha =2)$.}
\label{fig:var_num_compare}
\end{figure}

Figure \ref{fig:var_num_compare} shows the comparison of the single- and
three-peak VA solutions to the numerical one, obtained by means of the
imaginary-time propagation method (ITP) \cite{ITP1,ITP2}, for the
non-interacting condensate ($g=0$). Good agreement is observed between the
three-peak VA solution and the numerical one.

%%%%%%%%
%We compapre numerical and variations results both for positive and negative g in Figs. 3 and 4 of the revised manusrcipt.
%%%%%%%%
%{\LARGE [It would be very
%relevant to display a similar comparison for %}$g\lessgtr 0${\LARGE .]}

\begin{figure}[htb]
\includegraphics[width=0.4\textwidth]{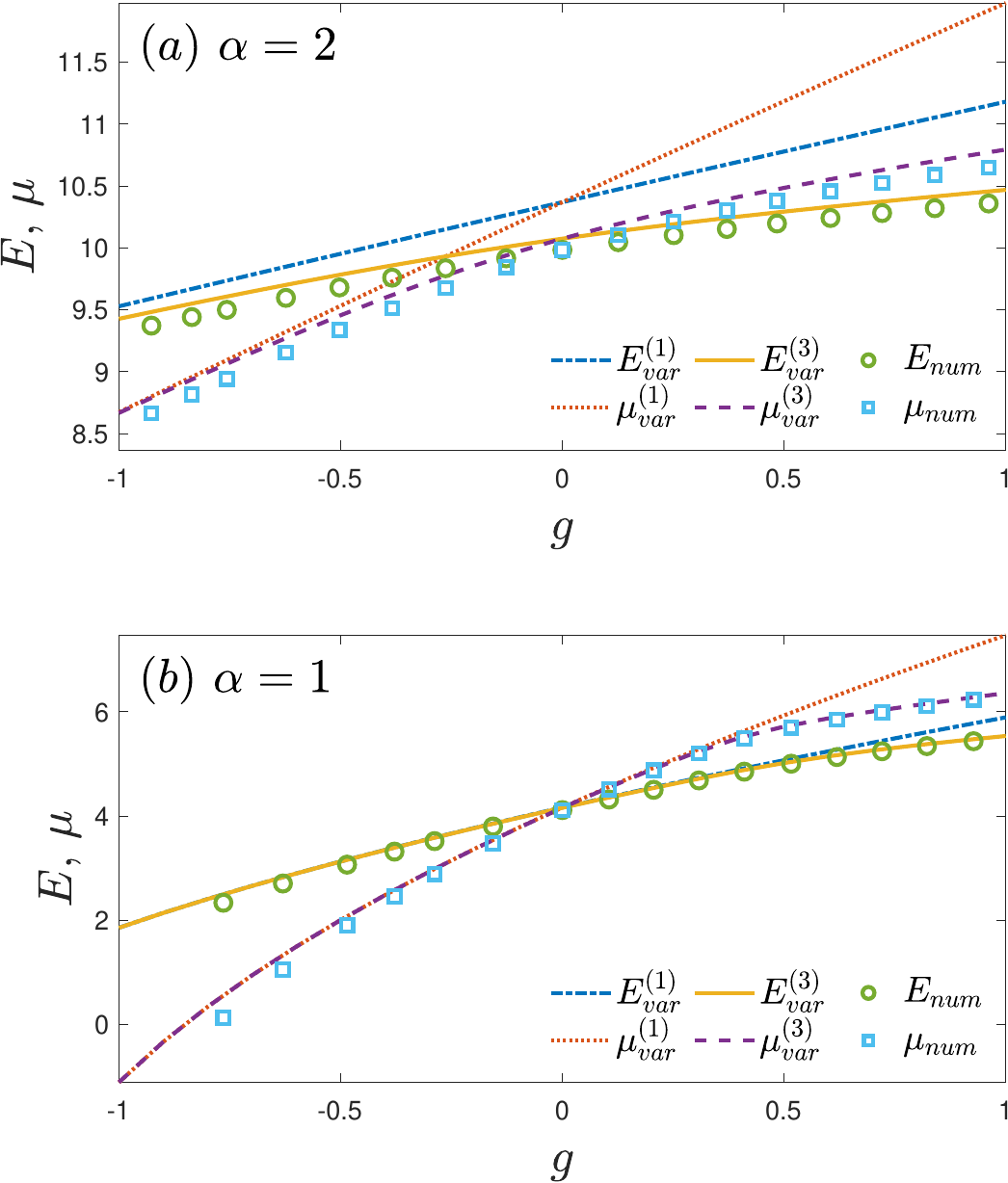}
\caption{Numerical results (circles and squares) and their VA-produced counterparts,
obtained by means of the ansatz admitting the single or three peaks, for the dependence of energy $E$ and
chemical potential $\protect\mu $ on the coupling constant $g$, for
different values of LI: (a) $\protect\alpha =2$ and (b) $\protect\alpha =1$.
Solid and dashed lines represent the energy and chemical potential, respectively, of the three-peak VA solutions, while dash-dotted and dotted lines correspond to the energy and chemical potential, respectively, of the one-peak VA solutions.}
\label{fig:E_mu}
\end{figure}

The consistency between numerical and three-peak variational solutions is further illustrated in Fig. \ref{fig:E_mu} using such integral characteristics as energy ($E$) and chemical potential ($\mu$).
It is evident that for lower values of the Lévi index $\alpha$, the single-peak ansatz yields sufficient accuracy over a broad range of coupling constant values. This reflects the tendency of the wave function to localize on a single lattice site as the diffraction term is suppressed with decreasing $\alpha$.

%%%%%

%\vspace{0.5cm}

%It can be seen that for the lower value of %the L\'{e}vi index $\alpha$, one-peak %ansatz also provides sufficient accuracy in %a wide region of the coupling constant %values, which reflects the fact that the %wave function tends to be concentrated in a %single lattice site as diffraction term is %suppressed by reducing the L\'{e}vi index %$\alpha$.

\begin{figure}[htb]
\includegraphics[width=0.45\textwidth]{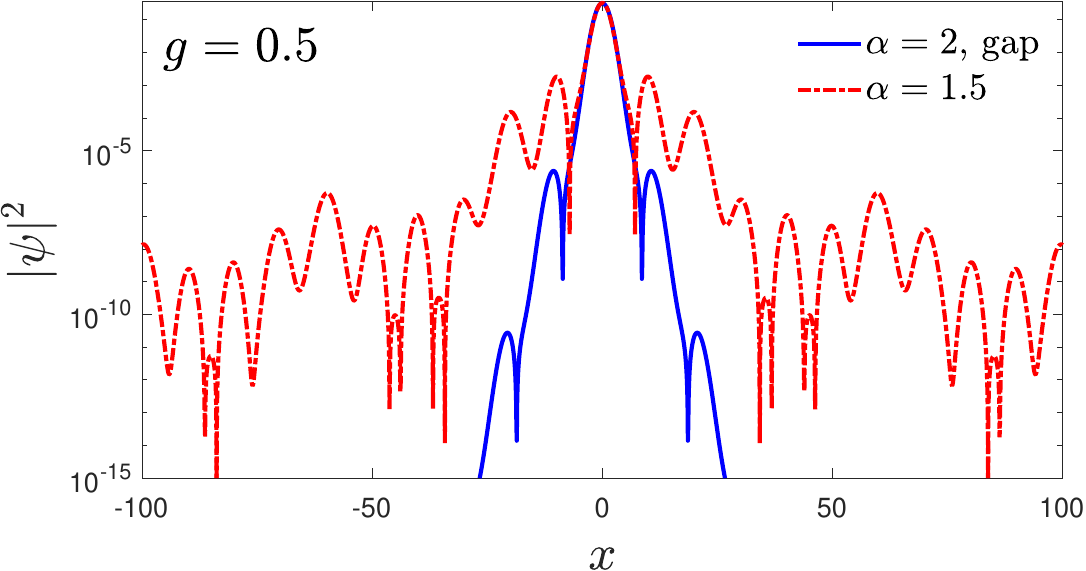}
\caption{\textcolor{black}{Comparison of the gap soliton supported by the periodic potential in the combination with the normal diffraction (the solid blue curve) 
and the localized state supported by the quasiperiodic potential in the combination with the fractional diffraction (the dashed-dotted red curve). 
The vertical axis represents the logarithmic scale.}
}
\label{fig:comparing_with_gap_solton}
\end{figure}

{\color{black}
\subsection{Gap solitons in periodic potential}

The soliton-like solutions presented in this work differ from traditional gap solitons. The gap solitons emerge in energy-spectrum gaps due to the 
action of the self-repulsive nonlinearity. In Fig. \ref{fig:comparing_with_gap_solton}, we compare these two species of the localized states. Note that the 
solution supported by the quasiperiodic potential in the combination with the fractional diffraction ($\alpha=1.5$, the dashed red curve) features prominent tails, 
in sharp contrast with the strongly localized shape of the gap soliton, which is supported by the periodic potential in the combination with the 
normal (non-fractional) diffraction ($\alpha=2$, the solid blue curve). Chemical potentials are $\mu\approx 0.33$ and $\mu\approx 0.36$ respectively. These numerical solutions, obtained by means of the Squared-Operator Iteration Method \cite{yang2010nonlinear}, align with prior results for the gap solitons \cite{Pelinovsky2004}.
In Ref. \cite{Pelinovsky2004}, the periodic potential is defined as:
\begin{equation}
    V_P(x)=V_0\sin^2\left(\frac{\pi x}{d}\right),\label{eq:Vp}
\end{equation}
where $V_0=1$ and $d=10$.
To construct a quasiperiodic potential, we introduce a second mode, yielding:
\begin{equation}
    V_{QP}(x)=\frac{V_0}{1+r}\left[\sin^2\left(\frac{\pi x}{d}\right)+r\sin^2\left(\frac{\pi x}{d'}\right)\right],
\end{equation}
where $r=0.2$, $d'/d=\varphi/2$, $\varphi\approx 1.618$ is the golden ratio. $V_{QP}$ is normalized to have the same amplitude as $V_P(x)$ in Eq. (\ref{eq:Vp}).

}
\clearpage
\twocolumngrid

\end{document}